%% file: main.tex
\documentclass[journal]{IEEEtran}

\ifCLASSINFOpdf
\else

\fi

\usepackage{amsmath}
\usepackage{amsfonts}
\usepackage{circuitikz}
\usepackage{algorithm}
\usepackage{algpseudocode}
\usepackage[
backend=biber,
style=ieee,
sorting=none
]{biblatex}
\usepackage{siunitx}
\usepackage{booktabs}
\usepackage{todonotes}
\usepackage{subcaption}
\usepackage{relsize}
\DeclareSIUnit[quantity-product = {}]
\amperehour{\text{Ah}}
\graphicspath{{Pics/}}

\newcommand{\ts}[1]{_{\mathrm{#1}}}
\newcommand{\tu}[1]{^{\mathrm{#1}}}
\newcommand{\mb}[1]{\mathbf{#1}}
\newcommand*\diff{\mathop{}\!\mathrm{d}}

\begin{document}
%
\title{Learning battery model parameter dynamics from data with recursive Gaussian process regression}
%
%
%

\author{Antti~Aitio, Dominik~Jöst, Dirk~Uwe~Sauer, David~A.~Howey,~\IEEEmembership{IEEE Senior Member} 
	\thanks{A.\ Aitio, D.\ Jöst are joint first authors.}
	\thanks{D.A.\ Howey and A.\ Aitio are with the Department
		of Engineering Science, University of Oxford}
	\thanks{D.\ Jöst and D.U. Sauer are with the Chair for Electrochemical Energy Conversion and Storage Systems at the Institute for Power Electronics and Electrical Drives (ISEA), RWTH Aachen University and Jülich Aachen Research Alliance, JARA-Energy}
	\thanks{D.U.\ Sauer is with Helmholtz Institute Münster (HI MS), IEK 12, Forschungszentrum Jülich and Institute for Power Generation and Storage Systems (PGS) at the E.ON ERC, RWTH Aachen University}}


\maketitle

\begin{abstract}
Estimating state of health is a critical function of a battery management system but remains challenging due to the variability of operating conditions and usage requirements of real applications. As a result, techniques based on fitting equivalent circuit models may exhibit inaccuracy at extremes of performance and over long-term ageing, or instability of parameter estimates. Pure data-driven techniques, on the other hand, suffer from lack of generality beyond their training dataset. In this paper, we propose a hybrid approach combining data- and model-driven techniques for battery health estimation. Specifically, we demonstrate a Bayesian data-driven method, Gaussian process regression, to estimate model parameters as functions of states, operating conditions, and lifetime. Computational efficiency is ensured through a recursive approach yielding a unified joint state-parameter estimator that learns parameter dynamics from data and is robust to gaps and varying operating conditions. Results show the efficacy of the method, on both simulated and measured data, including accurate estimates and forecasts of battery capacity and internal resistance. This opens up new opportunities to understand battery ageing in real applications.
\end{abstract}


\IEEEpeerreviewmaketitle

\section{Introduction}

Demand for battery systems is increasing rapidly as efforts to decarbonise electricity grids and electrify mobility gather pace \cite{Figgener2020TheReview}. Due to their long lifetime and high energy density, Li-ion cells have become the workhorse in battery systems \cite{IoannisTsiropoulos2018Li-ionGrowth}. Although the cost of these has dramatically decreased in the last decade \cite{Ziegler2021DeterminantsDecline}, the economics of storage needs to further improve to increase take-up, notably in applications where battery systems are not yet competitive in terms of levelized cost  \cite{DNVRenewablesAdvisory2021Techno-economicAfrica}. 
Also, given the risks of Li-ion cell demand outpacing the supply of the required raw materials \cite{IEA2021TheTransitions}, it is crucial that the performance of existing systems, especially in terms of lifetime, is maximised. A key element in improving the overall cost-effectiveness of Li-ion batteries is accurate estimation and prediction of battery state-of-health (SOH), which can improve lifetime, warranty and insurance costs, system safety and timing of maintenance. Accurate SOH estimation and prediction, especially using field data, opens up additional possibilities for second-life applications and helps greatly in `closing the loop' in terms of understanding the impact of design on real-world performance \cite{Sulzer2021TheData}.

While these issues have been extensively studied at cell level in laboratory environments, relatively little work has been done considering real-world usage data \cite{Sulzer2021TheData}. More complex challenges in SOH estimation using real-world data arise from the lack of controlled operating conditions, poorer (and often unknown) sensor accuracy, possible data gaps, and the lack of granularity of measurements when dealing with modules or packs as opposed to single cells. Also, when dealing with multiple cells simultaneously, cell-to-cell variability will further complicate estimation.

Battery SOH estimation methods are usually categorised into model-driven and data-driven approaches \cite{Berecibar2016}. The former consists of repeatedly fitting battery model parameters to input-output data, whereby the parameter estimates---such as resistance and capacity---reflect SOH, enabling tracking of SOH. Commonly this has been done using observers, including recursive state-parameter estimation techniques such as nonlinear approximations of the Kalman filter \cite{Baba2015,Plett2004,Plett2006}, and more rigorous methods from control theory that guarantee convergence of estimates via stability criteria \cite{Kim2010AObserver,Ascencio2019AugmentedBatteries}. 

Prognosis (i.e., future prediction) in this framework is achieved using a separate model for the evolution of parameters over battery lifetime, and this can range from a random walk \cite{Baba2015,Plett2004,Plett2006} to semi-empirical curve fits of trajectories that may be re-parameterised over lifetime using adaptive methods such as particle filtering \cite{Saha2009,Guha2018StateModels}, a Bayesian approach that also provides parameter uncertainty estimates. 
Model-driven approaches tend to use rather simple equivalent-circuit models because they have relatively few parameters that need to be fitted, whereas parameterising physics-based models, such as those within the Doyle-Fuller-Newman framework \cite{Doyle1993,Fuller1994}, is plagued by poor identifiability \cite{Aitio2020BayesianDynamics}. This is mainly due to a lack of reference electrodes in commercial cells which means that decoupling the positive and negative half-cell potentials is very difficult. In addition, physics-based models require a large set of parameters to be estimated or known a priori. However, using fixed-value equivalent circuit parameters to gauge battery SOH will give noisy estimates because parameters tend to vary as functions of battery internal states and operating conditions \cite{Gomez2011,Remmlinger2011,Aitio2021PredictingLearning}.

In contrast, data-driven methods for SOH diagnosis or prognosis attempt to map from operating data to SOH (usually defined in some consistent way, e.g.\ as the constant-current discharge capacity) either using the raw measurements directly \cite{Chaoui2017StateNetworks,Li2021OnlineNetworks} or via pre-defined features calculated from measurements \cite{Richardson2019,You2016,Greenbank2021AutomatedLife}. To obtain these nonlinear mappings, supervised machine-learning techniques such as neural networks \cite{Chaoui2017StateNetworks,Li2021OnlineNetworks}, Gaussian process regression  \cite{Richardson2019,RichardsonBatterymodel2019,Richardson2017} and relevance vector machines \cite{Zhang2017CapacityRVM} have been used. Depending on the choice of inputs, data-driven methods may be used to estimate either the present or the future SOH. If the inputs are chosen so that they consist of aggregated usage features to date (rather than current, voltage and temperature data from a single cycle), then these models can be used directly to forecast future SOH  \cite{Greenbank2021AutomatedLife}.

In this paper, we present a `hybrid' method of SOH diagnosis and prognosis that combines the model-driven and data-driven paradigms. Specifically, we parameterise a simple equivalent circuit model from experimental data using Gaussian process regression,  enabling us to describe the circuit parameters as smooth functions of time and operational conditions, rather than assume they are constant. This produces a more accurate circuit model because the underlying electrochemical processes are captured more realistically---for example, the reaction kinetics can be considered to be a nonlinear resistor that depends on current, temperature and state of charge as might be expected from  Butler-Volmer kinetics \cite{Brucker2021Grey-boxEquations}. To date, estimating battery equivalent circuit parameters as functions has received limited attention, although SOC-dependencies were recently investigated using linear parameter-varying models \cite{Fan2022State-of-chargeRegression}.

Using a computationally efficient implementation of Gaussian process regression \cite{ArnoSolin2016}, we show how battery states and state-dependent circuit parameters may be estimated simultaneously in an observer-like framework,  with computational effort scaling linearly with the number of rows in the input-output data.  The method is both battery-chemistry and construction agnostic, and the only prerequisite is a  lab measurement of the full cell open-circuit voltage as a function of state of charge. The  framework yields both a current estimate of SOH and a future prediction of SOH at any usage point with little extra computational effort.

\section{Gaussian process regression}

A Gaussian process (GP) is defined as a collection of random variables where any subset is jointly Gaussian-distributed \cite{Rasmussen2006}. Consequently, a GP defines a distribution over functions over an input $\mb{x}$, characterised by a mean and a covariance,
\begin{equation}
\begin{aligned}
    f(\mb{x}) &\sim \mathcal{GP}\left(m(\mb{x}),k(\mb{x},\mb{x}')\right), \\
    m(\mb{x}) &= \mathbb{E}[f(\mb{x})], \\
    k(\mb{x},\mb{x}') &= \mathbb{E}[\left(f(\mb{x})-m(\mb{x})\right)\left(f(\mb{x}')-m(\mb{x}')\right)],
\end{aligned}
    \label{eqn:GP_prior}
\end{equation}
where $k(\mb{x},\mb{x}')$ is the kernel function describing the covariance of the GP. Without loss of generality, we set the mean function $m(\mb{x})=0$. Given the definition of $f(\mb{x})$, the aim is to map inputs $\mb{x}$ to the outputs $\mb{y}$ such that
\begin{equation}
    \mb{y} = f(\mb{x}) + \epsilon ~,~ \epsilon \sim \mathcal{N}(0,\sigma_n^2).
    \label{eqn:GPreg}
\end{equation}
The fitting process consists of determining the posterior-predictive distribution, which can be used to predict values of $\mb{y}$ for any point in $\mb{x}$. Assuming zero-mean i.i.d.\ Gaussian measurement error $\epsilon$ in the observations $\mb{y}$, the posterior-predictive distribution is also Gaussian, with mean and covariance for a test point $\bf{x}_*\in\mathbb{R}\tu{d}$ given by %
\begin{subequations}
\begin{align}
    \mu(f_*) &= \mb{k}\ts{*,\bf{X}}\left[\mb{K}\ts{\mb{X}}+\sigma_n^2\mathbf{I}\right]^{-1}\mb{y} \label{eqn:p_mean} \\
    \text{cov}(f_*) &= \mb{k}\ts{**} - \mb{k}\ts{*,\mb{X}}\left[\mb{K}\ts{\mb{X}}+\sigma_n^2\mathbf{I}\right]^{-1}\mb{k}\ts{*,\mb{X}}\tu{T},
    \label{eqn:p_var}
\end{align}
\label{eqn:GPposterior}%
\end{subequations}
where $\mb{I}$ is the identity matrix of size $n$, where $n$ is the number of rows in the training data. For notational brevity we use $\mb{k}\ts{**}=k(\bf{x}_*,\bf{x}_*)$, $\mb{K}\ts{\bf{X}} = k(\bf{X},\bf{X})$ and $\mb{k}\ts{*,\bf{X}}=k(\bf{x}_*,\bf{X})$, and $\mb{X}\in\mathbb{R}\tu{n\times d}$ denotes the training input matrix.

For a zero-mean GP, the model is defined by the training data and the parameters of the kernel function $k(\bf{x},\bf{x}')$, known as `hyperparameters'---these describe smoothness, magnitude, periodicity and so on, depending on the chosen kernel  \cite{Rasmussen2006}. Using GP regression to fit data therefore also requires estimation of the hyperparameter vector $\theta$ given the training data. In the Bayesian framework, the posterior distribution of the hyperparameters $\theta$ given the training data is
\begin{equation}
\begin{aligned}
    p(\theta|\mb{X},\mb{y}) &=  \displaystyle\frac{p(\theta)\int p(\mb{y} | \theta,f,\mb{X}) p(f|\mb{X},\theta)\diff f }{p(\mb{y}|\mb{X})},
    \label{eqn:hyper_posterior}
\end{aligned}
\end{equation}
where $p(\theta)$ is the prior over the hyperparameters.
As the hyperparameter posterior is not tractable, it may be approximated using either the Laplace method \cite{Pietilainen2010ApproximationsProcesses}, Markov chain Monte-Carlo techniques, or variational inference \cite{Kucukelbir2017AutomaticInference}. If the full posterior can be approximated, then the hyperparameters may be marginalized (i.e.\ the whole distribution may be used) when making the GP prediction \cite{https://doi.org/10.1002/sim.3895}. However, for computational efficiency, we used maximum likelihood estimates of $\theta$ and these were obtained by setting up a uniform prior $p(\theta)$ and then maximising the logarithm of the numerator of  (\ref{eqn:hyper_posterior}) (i.e.\ the log marginal likelihood), given by \cite{Rasmussen2006}
\begin{multline}
    \log \int p(\mb{y} | \theta,f,\mb{X}) p(f|\mb{X},\theta)\diff f = \\
     -\frac{1}{2}\mb{y}\tu{T}\left[\mb{K}\ts{\mb{X}}-\sigma_n^2\mathbf{I}\right]^{-1}\mb{y} -\frac{1}{2}\log | \mb{K}\ts{\mb{X}} |- \frac{n}{2}\log 2\pi,
    \label{eqn:NLML}
\end{multline}
where $|\mb{K}\ts{\mb{X}}|$ denotes the  determinant of $\mb{K}\ts{\mb{X}}$.

Unfortunately, the standard approach for estimating hyperparameters or making model predictions with GP regression suffers from the so-called `big-n' problem, i.e.\ poor computational scaling, because both (\ref{eqn:GPposterior}) and (\ref{eqn:NLML}) require the inversion of an $n\times n$ matrix. This usually scales computationally as $\mathcal{O}(n^3)$ and can become numerically unstable for larger matrices. Several solutions for this have been proposed, such as sparse GP regression \cite{Quinonero-Candela2005}, structured approaches \cite{Saat2011}, reduced-rank methods \cite{Solin2019}, or GPU parallelisation \cite{Gardner2018GPyTorch:Acceleration}. For the purposes of this study, we implemented a recursive method \cite{ArnoSolin2016} that enabled a unified framework for simultaneous estimation of battery states and parameters. This scales as $\mathcal{O}(n)$ with the number of data rows, making it an efficient option for time series data.

\subsection{Recursive GP regression}

Recursive estimation of the GP posterior-predictive distribution (\ref{eqn:GPposterior}) and log marginal likelihood (\ref{eqn:NLML}) may be achieved by interpreting a Gaussian process as the solution to a linear time-invariant stochastic (partial) differential equation \cite{Lindgren2011}. This means that a GP may be represented by a spatio-temporal linear dynamic system \cite{ArnoSolin2016} of the type 
\begin{subequations}
\begin{align}
    \frac{\partial f(\mb{x},t)}{\partial t} &= \mathcal{F} f(\mb{x},t) + \mathbf{L}\omega(\mb{x},t) \\
    \mb{y}\ts{t} &= \mathcal{H}\ts{t}f(\mb{x},t) + \epsilon\ts{t}~,~\epsilon\sim\mathcal{N}(0,\sigma_n^2),
\end{align}\label{eqn:LTI_SDE}%
\end{subequations}
where $f(\mb{x},t)$ represents the GP at `position' $\bf{x}$ and time $t$, $\mathcal{F}$ and $\mathcal{H}\ts{t}$ are linear operators, $\mb{L}$ is a dispersion matrix, and $\omega(\mb{x},t)$ is spatially resolved white noise. The observation noise term $\epsilon\ts{t}$ is the same as in the standard (`batch') GP regression formulation (\ref{eqn:GPreg}). If the kernel function is separable, so that
\begin{equation}
  k\left((\mathbf{x},t),(\mathbf{x}',t')\right) = k(\mb{x},\mb{x}')k(t,t'),
\end{equation}
then the linear operator $\mathcal{F}$, becomes a constant matrix (denoted $\mb{F}$). In this study, all kernel formulations over the input space $(\mb{x},t)$ are separable. The link between the dynamic system and the kernel function describing the GP means that the kernel function may be directly mapped to $\mb{F}$, $\mathcal{H}\ts{t}$ and the spectral density of the white noise process $\omega(\mb{x},t)$. The details of this may be found in S\"arkk\"a et al.\ \cite{Sarkka2013b,Sarkka2012a} and Solin \cite{ArnoSolin2016}.

After formulating the GP as the solution of a linear dynamic system, estimation of the posterior-predictive distribution and hyperparameter log marginal likelihood can be achieved recursively using a Kalman filter and Rauch-Tung-Striebel smoother \cite{Rauch1965}. In the following sections, we describe a method for implementing this in the context of battery modelling, alongside  state estimation, allowing us to estimate model parameter dependencies on operating conditions and states in a computationally efficient way.

\section{Combining circuit models and GP regression}\label{sec:joint_estim}

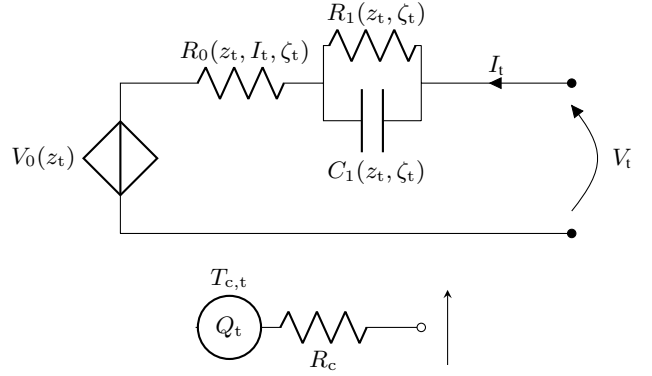
\begin{figure}
\small
\begin{circuitikz}
\draw (0,0) to[european controlled voltage source,l=\ $V_0(z\ts{t})$] (0,2) -- (0.5,2)
    to[R,l,l=\ {$R_0(z\ts{t},I\ts{t},\zeta\ts{t})$}] (2.7,2)
    (2.7,1.5) -- (2.7,2.5)
    to[R=\ {$R_1(z\ts{t},\zeta\ts{t})$}] (4,2.5)
    (2.7,2.5) -- (2.7,1.5)
    to[C,l_=\ {$C_1(z\ts{t},\zeta\ts{t})$}] (4,1.5)
    (4,2.5) -- ++ (0,-1)
    (4,2) to [short, -*, i<=$I_\text{t}$] ++ (2,0) coordinate (s1)
    (0,0) to [short, -*] ++ (6,0) coordinate  (s2)
    (s2) to [open, v_>=$V_\textrm{t}$ ] (s1)
    (1.,-1.25) to [esource,l=$T\ts{c,t}$] (1.9,-1.25)
    (1.45,-1.25) node[]{$Q\ts{t}$}
    (1.9,-1.25)  to [R, l_=$R\ts{c}$] (3.5,-1.25)
    (3.5,-1.25) to [short, -o] (4.,-1.25)
    (4.25,-1.25) node[right=10] {$T\ts{amb,t}$};
    \draw [stealth-](4.35,-.75) -- (4.35,-1.75);
\end{circuitikz}
\caption{Li-ion model with first-order RC electrical circuit and lumped thermal model. All electrical parameters are modelled as Gaussian processes.}
\label{fig:RC_circuit}
\end{figure}

The approach we use to model a Li-ion cell is shown in Fig.\ \ref{fig:RC_circuit}. It consists of a first-order resistor-capacitor (RC) electrical circuit with a coupled  lumped thermal model. All four electrical parameters are considered to be GPs over state of charge and/or applied current, as well as lifetime, denoted $\zeta\ts{t}$. The thermal model consists of heat generation due to the total overpotential (i.e.\ voltages across the series resistor and parallel RC pair) and convection to the ambient environment, making the assumption that heat conduction through the cell is fast (i.e.\ the Biot number is small), such that the cell internal temperature is relatively uniform. Entropic heating was ignored.

The (nonlinear) continuous-time dynamics of the 3-state electro-thermal model of Fig.\ \ref{fig:RC_circuit} in state-space form are
\begin{equation}
\begin{aligned}
    \frac{\diff \mb{x}\ts{t}}{\diff t} = f\ts{B}(\mb{x}\ts{t},I\ts{t},\theta) \\
    \mb{y}\ts{t} = h(\mb{x}\ts{t},I\ts{t},\theta),
    \label{eqn:lib_statespace}
\end{aligned}
\end{equation}
with the dynamics $f\ts{B}$ for the battery state vector $\mb{x}\ts{t} = [z\ts{t}~V\ts{1,t}~T\ts{t}]\tu{T}$ given by

\begin{equation}
\begin{aligned}
    \frac{\diff z\ts{t}}{\diff t} & = I\ts{t}Q^{-1}(\zeta\ts{t}) \\
    \frac{\diff V\ts{1,t}}{\diff t} & = -\alpha(z\ts{t},\zeta\ts{t})V\ts{1,t}+\beta(z\ts{t},\zeta\ts{t})I\ts{t} \\
    \frac{\diff T\ts{c,t}}{\diff t}C\ts{c} & = -\frac{T\ts{c,t}-T\ts{amb,t}}{R\ts{c}} + V\ts{1,t}I\ts{t} + R\ts{0}(z\ts{t},I\ts{t},\zeta\ts{t})I^2\ts{t},
\end{aligned}
    \label{eqn:batt_dyn}
\end{equation}
where $z\ts{t}$ is the state of charge, $I\ts{t}$ the applied current (positive for charging) and $Q^{-1}(\zeta\ts{t})$ the inverse battery capacity as a function of lifetime $\zeta\ts{t}$. Lifetime $\zeta\ts{t}$ can be measured by calendar age or by total charge throughput. The voltage $V\ts{1,t}$ is across the RC pair, and its time dynamics are controlled by the functions $\alpha(z\ts{t},\zeta\ts{t})$ and 
$\beta(z\ts{t},\zeta\ts{t})$. These are related to the circuit parameters, via $\alpha=1/R\ts{1}C\ts{1}$ and $\beta=1/C\ts{1}$. The thermal model is parameterised by heat capacity $C\ts{c}$ and thermal resistance $R\ts{c}$, which are considered known. Given these dynamics, the outputs $\mb{y}\ts{t} = [V\ts{t}~T\ts{t}]\tu{T}$ are cell terminal voltage and temperature,
\begin{equation}
\begin{aligned}
    V\ts{t} & = V\ts{0}(z\ts{t}) + V\ts{1,t} + R\ts{0}(z\ts{t},I\ts{t},\zeta\ts{t})I\ts{t} \\
    T\ts{t} & = T\ts{c,t}
\end{aligned}
\label{eqn:outputs}
\end{equation}

The model could be extended to include temperature dependencies for the circuit elements, but as the experimental data used here (see Section \ref{sec:exp}) only has a small temperature range, of approximately \SI{5}{\celsius}, this was not necessary.
The four functions $Q^{-1}(\zeta\ts{t}),~\alpha(z\ts{t},\zeta\ts{t}),~\beta(z\ts{t},\zeta\ts{t})$ and $R\ts{0}(z\ts{t},I\ts{t},\zeta\ts{t})$ are all assumed to be affine transformations of independent zero-mean Gaussian processes, so that
\begin{equation}
    f \sim t\ts{f}\left(\mathcal{GP}(0,k\ts{f}(x,x'))\right) ~,~ x = [z~I~\zeta]~,~f\in\{Q^{-1},\alpha,\beta,R\ts{0}\}.
\end{equation}
The affine transformation in each case is 
\begin{equation}
    t\ts{f}(x) = c\ts{f}(1+x),
    \label{eqn:aff_trans}
\end{equation}
where $c\ts{f}$ is a constant. As each GP describing $Q\tu{-1}$, $\alpha$, $\beta$ and $R\ts{0}$ has a zero mean, setting $c\ts{f}$ effectively sets a nonzero prior mean for each circuit parameter---hence $c\ts{t}$ should be chosen so that it reflects the prior expectation of where the parameter lies. The reason for the transformation is to scale the system so that the GPs are in the unit range (making hyperparameter initialisation simpler) and to improve the numerical stability of the system dynamics (\ref{eqn:batt_dyn}) at the prior mean of the GP. The four functions describing the circuit parameters are also Gaussian processes because a Gaussian distribution remains Gaussian under arbitrary affine transformations.

The kernel function $k\ts{f}$ is constructed so that the time input ($\zeta\ts{t}$) is treated differently from the inputs consisting of the instantaneous operating conditions ($z\ts{t}$ and $I\ts{t}$). In this case, a non-stationary kernel function describes each of the four Gaussian processes in the time dimension, which allows for better extrapolation than a stationary kernel since the latter reverts back to the mean upon long-range extrapolation. The non-stationary kernel here is the Wiener velocity (WV) kernel, given by
\begin{equation}
k\ts{WV}(\zeta,\zeta') = \sigma^2\ts{\zeta} \left(\frac{\min^3(\zeta,\zeta')}{3}+|\zeta-\zeta'|\frac{\min^2(\zeta,\zeta')}{2} \right).
\label{eqn:nonstat-kernel}
\end{equation}
The kernel describing the process over state of charge $z\ts{t}$ and applied current $I\ts{t}$ is the squared exponential (SE) kernel,
\begin{equation}
k\ts{SE}(\mb{x},\mb{x}') = \sigma\ts{\mb{x}}^2\exp\left(-\frac{1}{2}\sum_\mb{x} \gamma\ts{x}(\mb{x}-\mb{x}')^2\right)~,~\mb{x} = [z~I]\tu{T},
\label{eqn:stat_kernel}
\end{equation}
where $\gamma\ts{x}$ are the inverse length scales of the inputs $\mb{x}$. 
The kernels (\ref{eqn:nonstat-kernel}) and (\ref{eqn:stat_kernel}) are combined by multiplying the two functions. In addition, a kernel describing short-term fluctuations over time in the parameters is added---this adds equally to all points in $\mb{x}$, so that the overall kernel is therefore
\begin{equation}
     k\ts{f} = \underbrace{k\ts{WV}(\zeta,\zeta')k\ts{SE}(\mb{x},\mb{x}')}_{\text{spatially resolved, smooth}} + \underbrace{k\ts{E}(\zeta,\zeta')}_{\text{noise}}
     \label{eqn:mult_k},
\end{equation}
where $k\ts{E}$ is the exponential kernel,
\begin{equation}
    k\ts{E}(\zeta,\zeta') = \sigma^2\ts{\zeta,r}\exp\left(-\gamma\ts{\zeta,r}|\zeta-\zeta'|\right),
    \label{eqn:E_kernel}
\end{equation}
where $\sigma\ts{\zeta,r}$, $\gamma\ts{\zeta,r}$ are the magnitude and inverse length scale of the noise process. Hence the evolution of parameters is decomposed into a longer-term smooth component and shorter-term fluctuations. When extrapolated, the latter decays quickly, so the extrapolation is smooth. Estimating the hyperparameter $\gamma\ts{\zeta,r}$ from the data gives an estimate of the autocorrelation of the noise process. If $\gamma\ts{\zeta,r}$ is high, then $k\ts{E}$ effectively describes white noise over time.

\section{Joint estimation of battery states and GPs}

\subsection{Discretisation and joint state vector}

To construct a finite-dimensional state-space representation of the GP, Eqns.\ (\ref{eqn:LTI_SDE}), the input space $\bf{x}$ for each GP $f\in {R\ts{0},\alpha,\beta} $ has to be discretised.  To this end, we chose $n\ts{z}$ evenly spaced discretisation points over SOC ($z$) to represent the GP for $\alpha, \beta$, which are only functions of SOC, and $n\ts{zI}$ points for $R\ts{0}$, which is a function of both SOC and applied current. In other words, $\alpha$, $\beta$ and $R\ts{0}$ are each represented by a state vector where each element corresponds to a point at a specific SOC and/or $I$. For $R\ts{0}$ the vector is effectively a stacked set of values at sampling points that represent a grid over discrete SOC and current values. Therefore the state vectors for the three GPs can be written as
\begin{equation}
\begin{aligned}
    \mb{x}\ts{\alpha} & = \begin{bmatrix} \alpha\ts{z_1} & \alpha\ts{z_2} & \hdots &\alpha\ts{z_{nz}} \end{bmatrix}, \\
    \mb{x}\ts{\beta} & = \begin{bmatrix} \beta\ts{z_1} & \beta\ts{z_2} & \hdots & \beta\ts{z_{nz}} \end{bmatrix}, \\
    \mb{x}\ts{R_0} & = \begin{bmatrix} R\ts{0,(z_1,I_1)} & R\ts{0,(z_2,I_2)} & \hdots & R\ts{0,(z_{nzI},I_{nzI})}
    \end{bmatrix}.
    \label{eqn:discrete_points}
\end{aligned}
\end{equation}
A joint state vector may then be constructed to estimate the states of the GP and the battery model simultaneously given the $n\ts{z}$ and $n\ts{zI}$ points in the input space for the respective GPs. Let the vector $\mathbf{x}\ts{Batt,t}$ denote the mean estimates of the battery states at time $t$,
\begin{equation}
    \mathbf{x}\ts{Batt,t} = \begin{bmatrix}
        z\ts{t} & V\ts{1,t} & T\ts{c,t} 
        \end{bmatrix}\tu{T},
\end{equation}
and the vector $\mathbf{x}\ts{GP,s,t}$ denote the state vector associated with the mean of each of the GPs describing the model parameters, so that
\begin{equation}
    \mathbf{x}\ts{GP,s,t} = \begin{bmatrix}
        \mb{x}\ts{Q^{-1},t} & \bf{x}\ts{\alpha,t} & \mb{x}\ts{\beta,t} & \mb{x}\ts{R_0,t}
        \end{bmatrix}\tu{T},
\end{equation}
where $\mb{x}\ts{Q^{-1},t}\in\mathbb{R}^2$, $\mb{x}\ts{\alpha,t}\in\mathbb{R}^{2n\ts{z}}$, $\mb{x}\ts{\beta,t}\in\mathbb{R}^{2n\ts{z}}$, $\mb{x}\ts{R_0,t}\in\mathbb{R}^{2n\ts{zI}}$. The dimensionality of the state vectors is due to the number of points used for the discretisation of $\mb{x}$ in each case and the order of the Markov process due to the Wiener velocity kernel $k\ts{WV}(\zeta,\zeta')$. Specifically, the Wiener velocity kernel (\ref{eqn:nonstat-kernel}) has a dynamic representation  \cite{ArnoSolin2016} as follows,
\begin{equation}
    \frac{\diff}{\diff t}\begin{bmatrix} x \\ \frac{\diff x}{\diff t} \end{bmatrix} =
    \begin{bmatrix}
    0 & 1 \\ 0 & 0
    \end{bmatrix}\begin{bmatrix} x \\ \frac{\diff x}{\diff t} \end{bmatrix} + \begin{bmatrix}
    0 \\ 1
    \end{bmatrix}\omega(t),
    \label{eqn:WV_dynamics}
\end{equation}
where the spectral density of noise $\omega(t)$ is a function of $\sigma\ts{\zeta,s}$. This means that the kernel multiplication (\ref{eqn:mult_k}) results in each GP being represented by both the current state and its first-order time derivative at each spatial location in $\mb{x}$.  The stationary exponential kernel (\ref{eqn:E_kernel}), describing the short length scale noise process, has a single state representation, 
\begin{equation}
    \frac{\diff x}{\diff t} = -\gamma\ts{\zeta,r}x + \omega(t),
\end{equation}
where the spectral density of the noise process $\omega(t)$ is a function of $\sigma\ts{\zeta,r}$. In this case, the GP state vector for $k\ts{E}$ is 
\begin{equation}
    \mb{x}\ts{GP,r,t} = [x\ts{Q^{-1},1,t}~~x\ts{\alpha,1,t}~~x\ts{\beta,1,t}~~x\ts{R\ts{0},1,t}].
\end{equation}
The overall joint state-parameter system `state' representation is then given by the concatenation of the battery and parameter (GP) state vectors, 
\begin{equation}
 \mathbf{x}\ts{t} = \begin{bmatrix}
      \mathbf{x}\ts{Batt,t} \\ \mathbf{x}\ts{GP,s,t} \\ \mathbf{x}\ts{GP,r,t} \end{bmatrix}.
\end{equation}

\subsection{Initialisation and solution}

The joint system is nonlinear and may be solved through time using an appropriate Bayesian filter. For computational speed, the 
extended Kalman filter is applied here. It is possible that other variants, such as the unscented Kalman filter or particle filter, might provide more accurate results, but the EKF was considered adequate in initial tests using simulated data (Section \ref{sec:simulation}). There are two timescales involved---the first is given by the sampling frequency of current, voltage and temperature data during cycling, which in this work is \SI{1}{\hertz}, and the second is related to the `extent of degradation', $\zeta$, which covers the lifetime of the battery and may be measured by cumulative charge throughput or a similar metric. 

\subsubsection{Initialisation}

The initialisation of the two subsystems (i.e., for the battery states and parameter-GPs, respectively) is consistent with the two timescales. A zero-mean GP is used to model the circuit parameters (nested inside the affine transformation (\ref{eqn:aff_trans})), which means that the initial mean estimates of the GPs are set so that $\mathbf{x}\ts{GP,t} = \mb{0}$. The initial GP covariance matrix is block diagonal due to the assumption that all GPs are independent from each other, so that
\begin{equation}
\mb{P}\ts{GP,0} = \mathlarger{\oplus}\Big(\mathlarger{\oplus}\ts{f}\left(\mb{P}\ts{f,WV,0}\right), \mathlarger{\oplus}\ts{f}\left(  \mb{P}\ts{f,E,0}\right)\Big),
\end{equation}
where the direct sum operator $\oplus$ denotes the construction of a block diagonal matrix of its arguments, and $\mathrm{f} \in \{Q^{-1}, \alpha, \beta, R\ts{0}\}$ and $\mb{U}\ts{f}$. The initial covariance for each `smooth' (i.e.\ WV kernel) GP is given by the Kronecker relation
\begin{equation}
    \mathbf{P}\ts{f,WV,0} = k\ts{f,\mb{x}}(\mb{U}\ts{f},\mb{U}\ts{f}')\otimes \mb{P}\ts{\zeta_0,WV},
    \label{eqn:P_kron}
\end{equation}
where $\mb{U}\ts{f}$ are the coordinates of the discrete points chosen for each kernel function (i.e.\ the coordinates of \eqref{eqn:discrete_points}). This is a discrete representation of the initial covariance of the spatially resolved white noise process. 

The initial covariance for the Wiener velocity process, $\mb{P}\ts{\zeta_0,WV}$, is determined by hyperparameters. In the standard formulation, the WV kernel (\ref{eqn:nonstat-kernel}) has zero covariance at $\zeta=0$. However, the parameters of the circuit model are nonzero at the beginning of life, so a modification is required. To reconcile this, the WV kernel may be `truncated' by replacing $\zeta$ with $\zeta\ts{0}$, a nonzero value (see Appendix \ref{sec:WV_init}). This gives the initial WV covariance as
\begin{equation}
\mathbf{P}\ts{\zeta\ts{0},WV} = \sigma\ts{\zeta,s}^2 \begin{bmatrix}
          \frac{1}{3}\zeta\ts{0}^{3} & \frac{1}{2}\zeta\ts{0}^2\\
          \frac{1}{2}\zeta\ts{0}^2 & \zeta\ts{0}
          \label{eqn:WV_sol}
    \end{bmatrix}.
\end{equation}
For the noise process, 
\begin{equation}
    \mb{P}\ts{E,0} = \sigma\ts{\zeta,r}^2.
\end{equation}
The GP mean and covariance describing the parameters only have to be initialised once for each battery. The battery states, on the other hand, have to be re-initialised whenever there is a gap in telemetry data. For parameter estimation, not all data are required because battery degradation is slow compared to the sampling frequency. Therefore, only a smaller number of specific charge/discharge cycles from within the larger dataset need to be selected. In this work, the simulated and experimental datasets each begin with a rest period, so we initialised the mean vector for battery states as 
\begin{equation}
    \mb{x}\ts{Batt,0} = \begin{bmatrix}
       V\ts{0}^{-1}(V\ts{t}) \\ 0 \\ T\ts{amb,t}
    \end{bmatrix}
    \label{eqn:batt_m_init}
\end{equation}
at the start of each cycle. The state covariance was initialised with fixed values, so that
\begin{equation}
    \mb{P}\ts{Batt,0} = \begin{bmatrix}
       P\ts{z,0} & 0  & 0 \\
       0 & P\ts{V_1,0} & 0 \\
       0 & 0 & P\ts{T,0}
    \end{bmatrix},
    \label{eqn:batt_P_init}
\end{equation}
and the overall system covariance $\bf{P}\ts{0}$ is formed by the block diagonal combination,
\begin{equation}
    \mb{P}\ts{0} =\begin{bmatrix} \mb{P}\ts{Batt,0} & \mb{0} \\ \mb{0} & \mb{P}\ts{GP,0} \end{bmatrix}.
\end{equation}

\subsubsection{Solution in time}
The mean and covariance of the joint system are solved in discrete time by the extended Kalman filter \cite{Sarkka2013b} with additive noise,
\begin{equation}
\begin{aligned}
    \mathbf{x}^-\ts{t} &= g(\mathbf{x}\ts{t-1}, I\ts{t-1}, T\ts{t-1}) \\
    \mathbf{P}^-\ts{t} &= \mathbf{G}\ts{t-1}\mathbf{P}^+\ts{t-1}\mathbf{G}\ts{t-1}\tu{T} + \mathbf{Q}\ts{t-1} +\lambda\ts{G,t-1}
\end{aligned}
\label{eqn:KF_prop}
\end{equation}
where $g$ describes both the system and parameter evolution dynamics, $\mathbf{G}\ts{t}$ is the local Jacobian matrix of $g$ at $\mathbf{x}\ts{t}$, $\mathbf{Q}\ts{t}$ is the joint discrete-time process covariance and $\lambda\ts{G,t}$ is an additional variance term arising from the posterior predictive variance of the GP. In addition, the predictive variances of $\alpha$, $\beta$ and $R\ts{0}$ have to be corrected for the uncertainty in the input variable $z\ts{t}$. On both these points, see Appendix \ref{sec:extra_vars}).

\subsubsection{State and covariance propagation}

Given estimates of $\alpha(z\ts{t})$, $\beta(z\ts{t})$, and $R\ts{0}(z\ts{t},I\ts{t})$, discrete time propagation of battery dynamics is approximated by using a zero-order hold on the applied current and linearising $g$ with respect to the battery states at time $t$. The propagation of the GP states $\mb{x}\ts{GP,t}$ is independent of the battery states and is linear, with the state transition is given by 
\begin{equation}
     \mb{x}\ts{GP,t} =\exp\Big(\mathlarger{\oplus}(\mathlarger{\oplus}\ts{f}(\mb{I}\ts{f} \otimes \mb{F}\ts{WV}),\mb{F}\ts{E}) \Delta \zeta\Big) \mb{x}\ts{GP,t-1},
     \label{eqn:state_trans_GP}
\end{equation}
where $\mb{I}\ts{n}$ is the identity matrix of size $n$, equal to the number of discrete `spatial' points propagated through time for each GP, and
\begin{equation}
        \mb{F}\ts{WV} = \begin{bmatrix} 0 & 1 \\ 0 & 0 \end{bmatrix} ~,~ \mb{F}\ts{E} = -\gamma\ts{\zeta.r}\mb{I}\ts{4}.
\end{equation}
The variable $\Delta\zeta\ts{t}$ is the time step size in the GP `degradation' timescale, which is larger than the time step of the system dynamics. Within each discharge cycle this is assumed constant, therefore requiring only a single evaluation of the matrix exponential \eqref{eqn:state_trans_GP} for each discharge cycle. To improve computational efficiency in fitting the GPs $\alpha$, $\beta$, and $R\ts{0}$  over battery lifetime, specific discharge cycles are down-selected from the raw data at an appropriate rate (in this case, one in thirty). Therefore, $\Delta\zeta$ is  the cumulative charge throughput over a period of weeks rather than seconds. The discrete time process noise covariance matrix $\mathbf{Q}\ts{t}$ is block diagonal, where the values for the battery states are fixed and the values for the GP are determined by the kernel function hyperparameters. More specifically, the WV kernel and exponential kernels have discrete time process variances as a function of the step $\Delta\zeta$,
\begin{equation}
    \mathbf{Q}\ts{t} = \oplus(\mathbf{Q}\ts{Batt} ,\mathbf{Q}\ts{GP})~,~
    \mathbf{Q}\ts{Batt} = \oplus(q\ts{z},q\ts{V\ts{1}} ,q\ts{T\ts{c}}),
\end{equation}
and the block diagonal GP process covariance is 
\begin{multline}
\mb{Q}\ts{GP}(\Delta \zeta) =  \mathlarger{\oplus}\Big(\mathlarger{\oplus}\ts{f}\left(k\ts{f,\mb{x}}(\mb{U}\ts{f},\mb{U}\ts{f}')\otimes\mb{Q}\ts{f,WV}(\Delta \zeta)\right), \\
\mathlarger{\oplus}\ts{f}\left(  Q\ts{f,E}(\Delta \zeta)\right)\Big).
\end{multline}
where $\mb{U}\ts{f}$ are the values of the coordinates of the discretisation points for each GP \eqref{eqn:discrete_points}. The exponential process covariance is scalar for each of the functions $f$ and given by
\begin{equation}
    Q\ts{E}(\Delta\zeta) = \sigma^2\ts{\zeta,r}\left(1-\exp(-2\gamma\ts{r}\Delta\zeta)\right).
\end{equation}

\subsection{Observation model}

Following the forward propagation of the system by (\ref{eqn:KF_prop}), the predicted voltage and temperature are given by (\ref{eqn:outputs}), which involves re-evaluating $R\ts{0}(z\ts{t},I\ts{t})$ with the latest estimate of the state vector $\mb{x}\ts{R_0,t}$. The predictive equation  (\ref{eqn:outputs}) is nonlinear in SOC ($z\ts{t}$), due to the open-circuit potential $V\ts{0}$ and $R\ts{0}(z\ts{t},I\ts{t})$. Using the EKF, (\ref{eqn:outputs}) is locally linearised to give the observation Jacobian $\bf{H}\ts{t}$. The dependency of $R\ts{0}$ on $z\ts{t}$, which itself is a Gaussian random variable, can be calculated using the method  discussed in Appendix \ref{sec:z_unc}. %
In addition, the uncertainty in $R\ts{0}$ from GP extrapolation to the current operating point, $\lambda\ts{H,t}$ is also incorporated into the predictive distribution for the output, which is
\begin{equation}
    \begin{bmatrix}
    V\ts{t} \\ T\ts{t} 
    \end{bmatrix}\sim \mathcal{N}(h(\mathbf{x}^-\ts{t},I\ts{t}),\mb{S}\ts{t}),
\end{equation}
where the covariance $\mb{S}\ts{t}$ of the output is given by
\begin{equation}
    \mb{S}\ts{t} = \mb{H}\ts{t}\mb{P}\ts{t}^-\mb{H}\ts{t}\tu{T}+\mb{R}+\lambda\ts{H,t},
    \label{eqn:KF_output}
\end{equation}
where $h$ is given by (\ref{eqn:outputs}), $\mb{R}$ is the measurement noise covariance matrix that is estimated, and the calculation of $\lambda\ts{H,t}$ may be found in the Appendix \ref{sec:extra_vars}.

\subsection{Hyperparameter optimisation and smoothed posterior}

The battery model dynamics and outputs depend on the hyperparameters of the kernel functions $k\ts{\zeta}$ and $k\ts{\mb{x}}$, which consist of the magnitudes $\sigma\ts{\zeta}$, $\sigma\ts{\zeta,r}$, $\sigma\ts{\mb{x}}$ and (inverse) length scales $\gamma\ts{\mb{x}}$. In addition, we estimate from the data the noise parameters $\sigma\ts{n,V}$ and $\sigma\ts{n,T}$ for the output voltage and temperature.  In order to find maximum likelihood estimates for the hyperparameters, the EKF recursion algorithm may be augmented to also update the negative log marginal likelihood (NLML) (\ref{eqn:NLML}) over the observations \cite{Sarkka2013b}, as shown in the last step of Algorithm \ref{tab:KF_full}. Calculating the NLML enables us to then optimize the hyperparameters by using a gradient based optimiser. Analytical solutions for the gradient of the NLML may also be calculated \cite{Mbalawata2013,Aitio2021PredictingLearning}, but in this case we applied automatic differentiation (AD) to calculate gradients.
\begin{algorithm}[t]
\small
\input{Tables/double_EKF_algo}
\caption{Extended Kalman filter algorithm with NLML calculation. States are initialised at the beginning of every cycle and the GP is propagated over the battery lifetime. The NLML is calculated recursively for all available data.}
\label{tab:KF_full}
\end{algorithm}

Once the hyperparameters have been estimated, the final step is to calculate a posterior distribution of the latent states $\mb{x}\ts{f,t}$ that is consistent with `batch-mode' GP regression. The forward filtering distribution $p(\mb{x}\ts{f,t}|\mb{y}\ts{1:t})$, Algorithm \ref{tab:KF_full}, only includes observations up until point $t$. However, we require the so-called smoothing distribution $p(\mb{x}\ts{f,t}|\mb{y}\ts{1:T})$, i.e., the marginal distributions of the latent states conditioned on \textit{all} available data. This is given by the RTS smoother \cite{Rauch1965}, which operates by applying a backward recursion through the filtering distribution calculated earlier. The smoothing distribution only needs to be calculated for the linear GP subsystem in the $\zeta$ timescale, therefore only requiring a number of steps equal to the number of discharge cycles in the data.

\begin{figure}[t]
    \centering
    \def\svgwidth{\columnwidth} 
    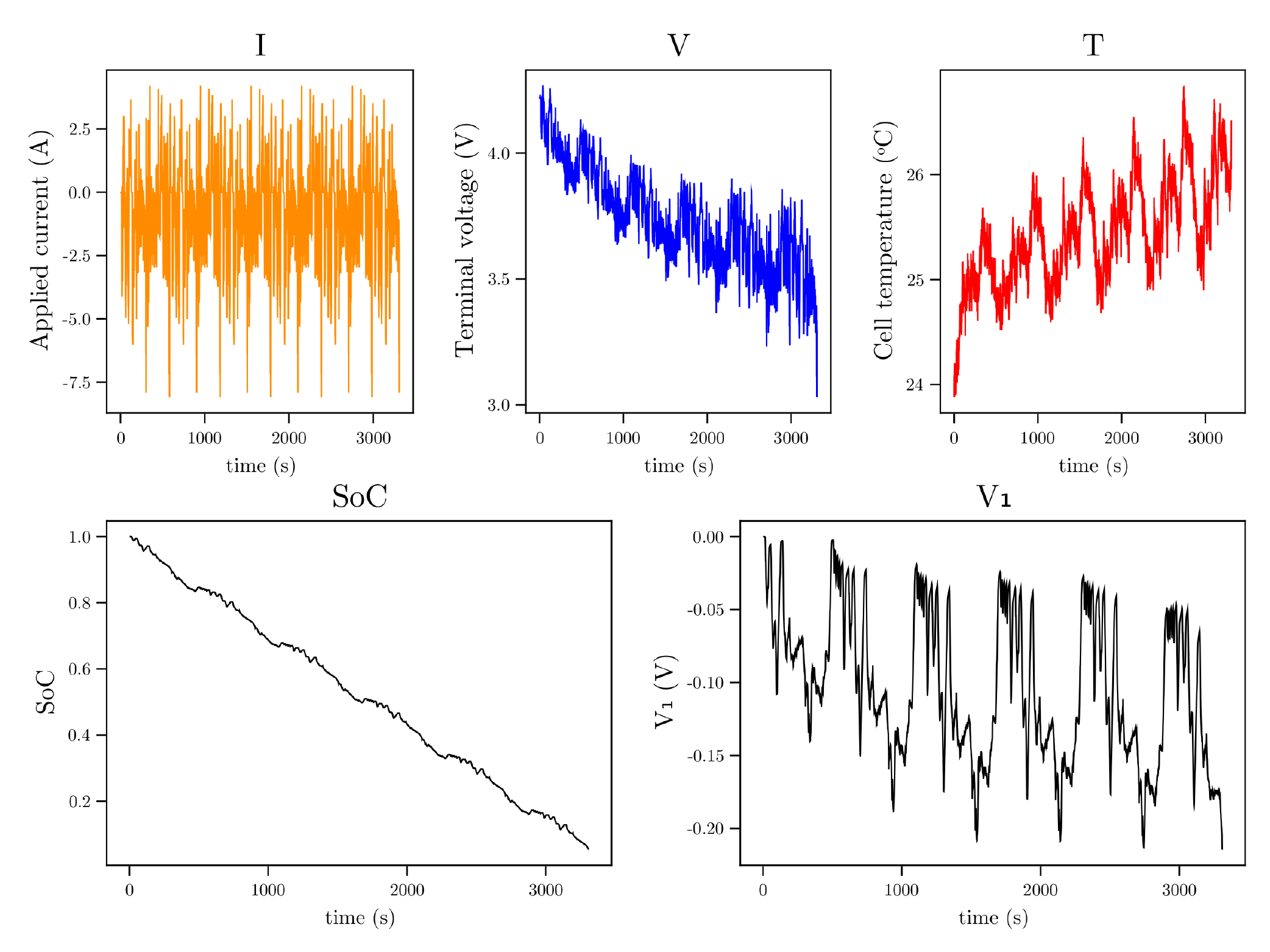
    \caption{Input/output data for simulation}
    \label{fig:sim_io}
\end{figure}

\section{Simulation results}\label{sec:simulation}

\begin{figure}[t]
    \centering
    \includegraphics[width=\columnwidth]{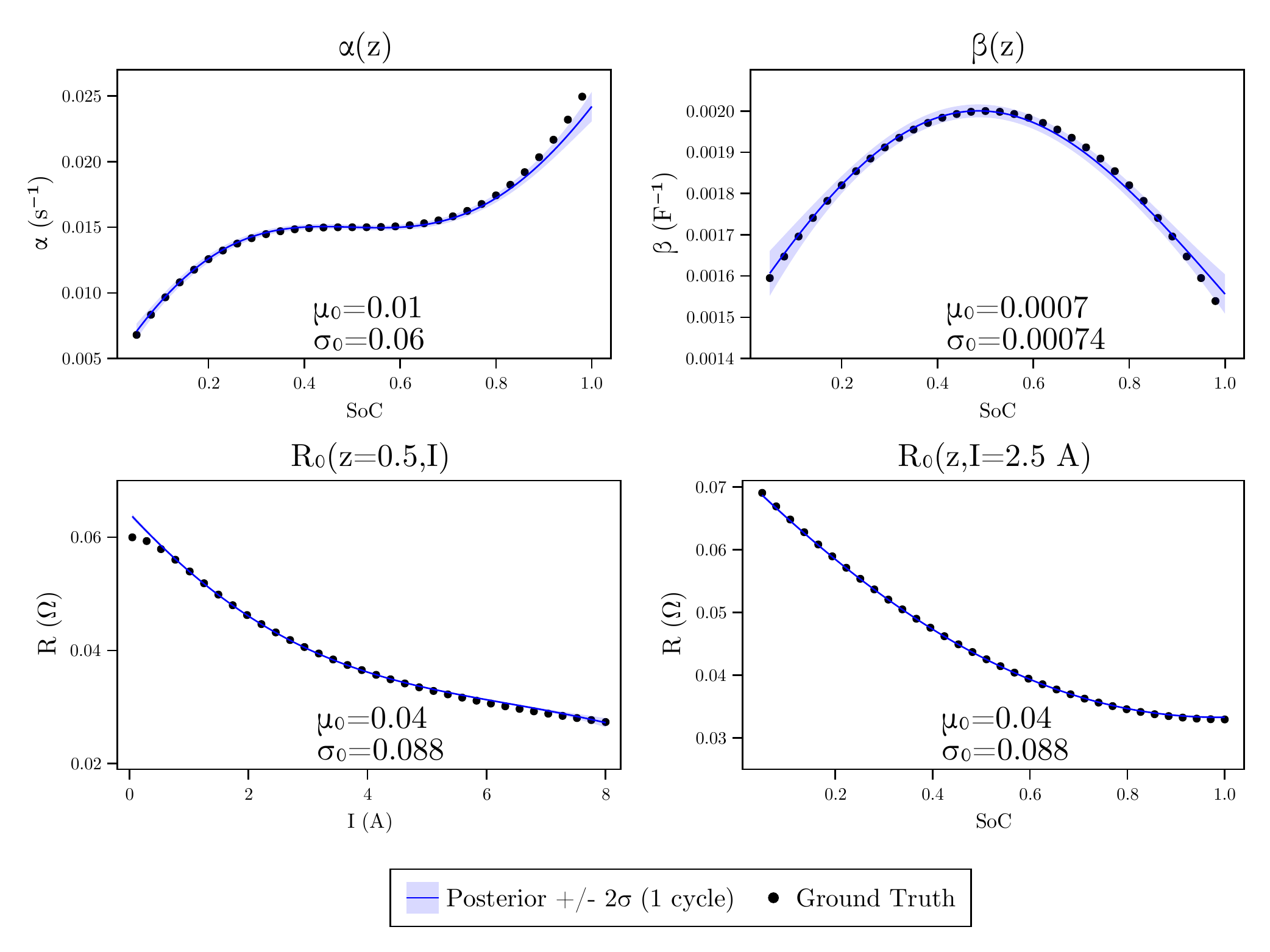}
    \caption{Ground truth functions for $\alpha(z)$, $\beta(z)$, $R\ts{0}(z,I)$ together with their GP estimates using 1 cycle of input/output data. Estimation errors are reported in Table \ref{tab:sim_params}.}
    \label{fig:sim_results}
\end{figure}

We now use data from simulations to obtain a known ground-truth and  demonstrate the effectiveness of the method proposed in this paper for identifying circuit parameters as functions of operating conditions. To this end, we simulated the voltage and temperature responses from the model (\ref{eqn:batt_dyn}) using a current profile from a US06 drive cycle \cite{GregoryL.Plett2015}. The ground-truth functions $\alpha,~\beta$ and $R\ts{0}$ were chosen arbitrarily, and are shown in Table \ref{tab:sim_params} alongside the other simulation parameters. To simulate measurement noise, zero-mean Gaussian noise was added to voltage and temperature measurements with standard deviations of \SI{5}{\milli\volt} and \SI{0.1}{\kelvin} respectively. The current profile, voltage and temperature responses, and the internal states of the model (i.e.\ $z$ and $V\ts{1}$) are shown in Fig.\ \ref{fig:sim_results}.
\begin{table*}[]
    \input{Tables/sim_parameters}
    \label{tab:sim_params}
\end{table*}
The joint GP/battery state estimator was applied to the simulated data, with 6 evenly spaced points over the range of state of charge $z$ used to describe the GPs $\alpha$ and $\beta$, and similarly an evenly spaced 4$\times$15 grid over observed values of $z,I$ was chosen for $R\ts{0}$. The constants in the transformation (\ref{eqn:aff_trans}) for $Q^{-1},~\alpha,~\beta$ and $R\ts{0}$ were set at \SI{1.09}{\per\amperehour}, \SI{0.01}{\per\second}, \SI{0.0007}{\per\farad} and \SI{0.04}{\ohm} respectively. The GP hyperparameters, consisting of the length scales for $\alpha, \beta, R\ts{0}$, the magnitudes for all four GPs, and the noise parameters $\sigma\ts{n,V}$ and $\sigma\ts{n,T}$, were estimated using a box-constrained Broyden-Fletcher-Goldfarb-Shanno (BFGS-B) optimisation algorithm implemented in the \texttt{Optim.jl} package in Julia \cite{mogensen2018optim}, using forward-mode automatic differentiation to calculate the NLML gradients \cite{RevelsLubinPapamarkou2016}. The constraints in the optimisation routine were imposed to guarantee numerical stability while optimising. For the purposes of identifiability, we only consider here the case where battery age $\zeta$ is constant, requiring therefore only a single discharge cycle in order to estimate the functions over operating conditions. %
In other words, the identifiability of the parameter functions can be considered independent of the input $\zeta$ in each case as long as the availability of data is similar throughout lifetime. If this is not the case, the GP framework will simply produce predictive posteriors with wider credible intervals where data is sparser. 

The results of the estimation are shown in Fig.\ \ref{fig:sim_results} and Table \ref{tab:sim_params}. In Fig.\ \ref{fig:sim_results}, projections of each function is shown.
From these, it is clear that the Gaussian process estimator accurately retrieves the ground-truth values for the circuit parameters. However, a small loss of accuracy occurs for $R\ts{0}$ at low current $I$ where the ground truth function has an inflection point. The maximum likelihood hyperparameter estimates are such that the GP assumes a long length scale for $R\ts{0}(z,I)$ over both input dimensions, giving a predictive posterior that extrapolates to a higher value as $I\rightarrow0$. This is partially due to numerical ill-conditioning at very low applied current i.e.\ a small error in voltage and/or current causes a large change in estimated $R\ts{0}$. The GP in this case relies on extrapolation from regions of higher $I$ where the function is more identifiable and smooth and has no turning points.

\section{GP Estimates from experimental data}\label{sec:exp}

The simulation work of the previous section shows that GP representations of known ground-truth functions for ECM parameters are able to be retrieved from input-output data. To apply the framework to real data, we cycled high energy 18650 Li-ion cells (Samsung SDI INR18650-35E, NCA vs.\ C+Si, \SI{3450}{mAh}) and measured current, voltage and temperature. The cells were mounted in a cell holder, cooled by active air-cooling in a temperature chamber (Binder MK240) and cycled with a battery tester (Digatron MCFT 20-5-60 ME), which has a datasheet accuracy of \mbox{$\pm$ \SI{40}{mA}/\SI{10}{mV}} and \mbox{$\pm$ \SI{20} {mA}/\SI{4}{mV}} after calibration. The measured time series data of current, voltage and temperature is available online  \cite{Jost2021Timeseries18650}. 

\begin{figure}
    \centering
    \includegraphics[width=\columnwidth]{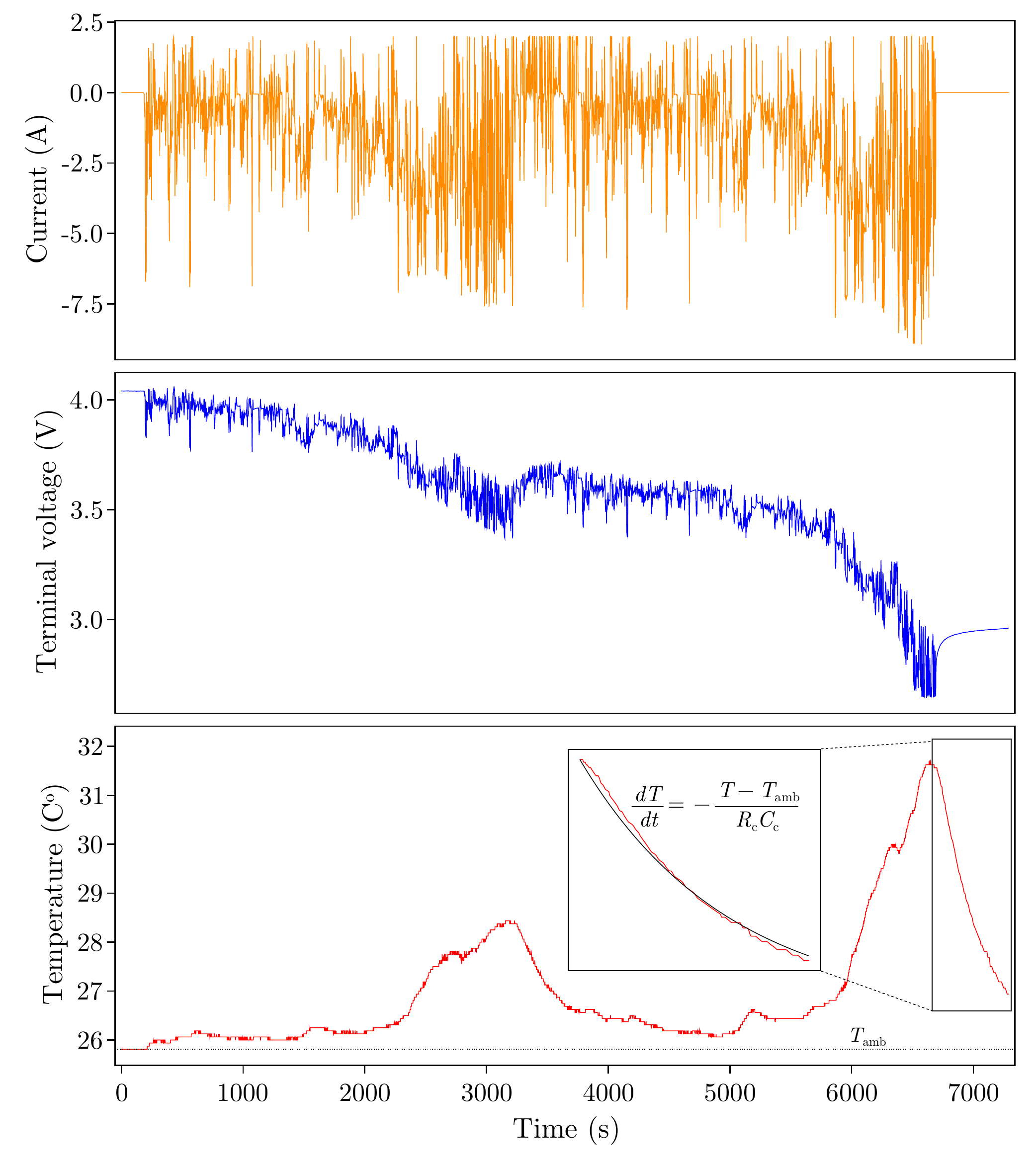}
    \caption{Current, voltage and temperature profiles for the drive cycle used in experimental set-up. To parameterise the thermal model, we estimated $R\ts{c}$ from the thermal relaxation at the end of the
    drive cycle using least-squares.}
    \label{fig:PowerProfile}
\end{figure}

The cycle ageing was conducted at a temperature of \SI{25}{\celsius} between $10 \%$ and $90 \%$ SOC. A constant-current constant-voltage (CCCV) charging protocol was used, with a current of 0.3C  and a cutoff-current of 0.02C, and discharging was achieved using a recorded and scaled drive cycle profile with an average current of approximately 0.4C, shown in Fig.\  \ref{fig:PowerProfile}. This cycling profile was recorded from a fully electric delivery van, and is composed of four sections. The first section, up to \SI{1135}{s}, represents inner-city driving. From there to \SI{2065}{s}, driving is on intercity roads. The following section up to \SI{2595}{s} was recorded on motorways. The final section was driven in a hilly region away from the motorway.

A checkup test procedure was conducted every 30 cycles, and consisted of 0.3C discharge capacity test after CCCV full charge followed by pulse tests at 3 SOC levels ($80\%,\ 50\%,\ 20\%$). The pseudo open-circuit voltage was determined at beginning of life using a full discharge at 0.02 C. To parameterise the thermal model in (\ref{eqn:batt_dyn}), the heat capacity of the cell (\SI{43.5}{\joule\per\kelvin}) was taken from literature \cite{Steinhardt2021Low-effortCells}. The thermal resistance $R\ts{c}$ was determined from the thermal relaxation following the first drive cycle at beginning of life, using a least-squares fit, illustrated in Fig.\ \ref{fig:PowerProfile}.

\subsection{GP Estimation}

We used the recursive framework described in Section \ref{sec:joint_estim} to estimate the electrical parameters in the model (\ref{eqn:batt_dyn}) for two example Li-ion cells (cell numbers 009 and 015) in the dataset. We assumed that ageing was negligible within consecutive drive cycles between check-up sequences, and therefore only the cycling data from the final drive cycle in each set of 30 repeats was used. Within each cycle, data were interpolated to a frequency of \SI{1}{\hertz} using the piece-wise cubic hermite method \cite{Carlson1980} to reduce the computational load for the EKF, avoiding the need to re-evaluate the matrix exponential in (\ref{eqn:state_trans_GP}). 

For independent validation of our method, we used the separate 0.3C capacity test and internal resistance data calculated as the average (over charge/discharge) $\Delta V/ \Delta I$ from the first second of the pulse tests. The dataset was split into two sections for each cell to investigate the method's ability to predict future degradation. Specifically, we left out the last 8 sets of cycles to be used in an out-of-sample setting to assess the ability of the GP to forecast future SOH evolution, giving an in-sample set of 19 and 20 discharge cycles for the two cells, respectively.

\subsubsection{Hyperparameter estimation}
\label{sec:hyper_est_detail}

The four GPs representing $Q^{-1}$, $\alpha$, $\beta$ and $R\ts{0}$ each have a multiplicative kernel function of the type (\ref{eqn:mult_k}), where in the case of $Q^{-1}$ the function $k\ts{x}(\mathbf{x},\mathbf{x}')$ is a constant because it has no dependency on $z$ or $I$. In total, the hyperparameter vector $\mathbf{\Theta}\ts{h}$ controlling the properties of the 4 GPs over the inputs $z$, $I$ and $\zeta$ contains 16 elements, so that
\begin{equation}
\begin{aligned}
    \mathbf{\Theta}\ts{h} &= [\mathbf{\Theta}\ts{\mathbf{SE,x}}~\mathbf{\Theta}\ts{WV,\zeta}~\mathbf{\Theta}\ts{E,\zeta}~\sigma\ts{n,V}~\sigma\ts{n,T}]~,~\text{where} \\
    \mathbf{\Theta}\ts{SE,\mathbf{x}} &= [\sigma\ts{Q^{-1}}~\sigma\ts{\alpha}~\sigma\ts{\beta}~\sigma\ts{R\ts{0}}~\gamma\ts{\gamma,z}~\gamma\ts{\beta,z}~\gamma\ts{R\ts{0},z}~\gamma\ts{R\ts{0},I}], \\
    \mathbf{\Theta}\ts{WV,\zeta} &= [\sigma\ts{Q^{-1},\zeta}~\sigma\ts{\alpha,\zeta}~\sigma\ts{\beta,\zeta}~\sigma\ts{R\ts{0},\zeta}], \\
    \mathbf{\Theta}\ts{E,\zeta} &= [\sigma\ts{\zeta,r}~\alpha\ts{\zeta,r}],
\end{aligned}
\end{equation}
where $\mathbf{\Theta}\ts{SE,\mathbf{x}}$ are the hyperparameters for the SE kernel and $\mathbf{\Theta}\ts{\zeta}$, $\mathbf{\Theta}\ts{WV,\zeta}$ are those for the WV and exponential kernels respectively, with measurement noise standard deviations given by $\sigma\ts{n,V}$ and  $\sigma\ts{n,T}$ for the terminal voltage and cell temperature respectively. We assumed that both cells share the same set of hyperparameters and all estimation was done using the summed NLMLs across the two cells in each scenario. We took several steps to reduce the computational effort required to estimate the hyperparameters $\mathbf{\Theta}\ts{h}$. 

Firstly, we assumed that the length scale and magnitude parameters for the GPs $\alpha$ and $\beta$ may be shared, as they both relate to the behaviour of the RC pair over SOC, hence $\sigma\ts{\alpha}=\sigma\ts{\beta}$ and $\gamma\ts{\alpha}=\gamma\ts{\beta}$. Secondly, we grouped the WV magnitude parameters $\mathbf{\Theta}\ts{WV,\zeta}$ into two, so that $\mathbf{\Theta}\ts{WV,\zeta} = [\sigma\ts{0,\zeta}~\sigma\ts{1,\zeta}]$, where $\sigma\ts{0,\zeta}$ was the WV kernel magnitude for $Q^{-1}$ and $\sigma\ts{1,\zeta}$ the WV magnitude for $\alpha$, $\beta$ and $R\ts{0}$. The rationale for this grouping arises from assuming that the degradation process, \textit{in relative terms} is similar for $\alpha$, $\beta$ and $R\ts{0}$, which are also dependent on operating conditions. This reduces the hyperparameter vector size to $\mathbf{\Theta}\ts{h}\in\mathbb{R}^{12}$.

Finally, to use all available data (288,639 rows of $I,V,T$ data in total for the in-sample set) to estimate  $\mathbf{\Theta}\ts{h}\in\mathbb{R}^{12}$ would still require substantial computational effort due to the high dimensionality of the optimisation problem. Therefore the hyperparameter estimation problem was split into two. As shown in Section \ref{sec:simulation}, the dependency of each of the functions $\alpha$, $\beta$ and $R\ts{0}$ on battery states and operating conditions may be inferred from a single cycle. Although this dependency may vary over the lifetime of the battery, we assumed that the hyperparameters controlling the GPs over $z$ and $I$ are constant throughout life. With this assumption, we used the first available cycle from the beginning of life to estimate the subset of hyperparameters $\mathbf{\Theta}\ts{\mathbf{x}}$. In addition to reducing the dimensionality of the estimation problem (as any hyperparameters related to $\zeta$ did not have to be estimated), the number of data rows required was reduced from 288,639 to 14,597. 

A multi-start process was used in the optimiser. First, we estimated the NLML using 1000 randomly chosen points for $\mathbf{\Theta}\ts{\mathbf{x}}$. From these, we then chose the 25 lowest NLML points and applied the same gradient-based optimization algorithm as in the simulation case, where the final $\mathbf{\Theta}\ts{\mathbf{x}}$ values chosen were those with the lowest overall NLML value. Following this,  $\mathbf{\Theta}\ts{\zeta}$ were estimated using maximum likelihood together with fixed $\mathbf{\Theta}\ts{\mathbf{x}}$ using the full in-sample dataset of 288,639 rows. In this case, the optimisation problem was only 4-dimensional so the number of iterations required was lower and only a single starting point was used, found by grid search. Again the box-constained BFGS algorithm was applied with automatic differentiation. The box-constraints additionally imposed a minimum on $\sigma\ts{\zeta,r}$ to guarantee smoothness of the WV process. The battery model process noise covariance matrix was fixed,
\begin{equation}
    \mathbf{Q}\ts{Batt} = \begin{bmatrix}
    10^{-12} & 0 & 0 \\ 0 & 10^{-6} & 0 \\ 0 & 0 & 10^{-4}
    \end{bmatrix}.
\end{equation}

\subsubsection{GP Posterior estimation and validation}

Given maximum likelihood estimates for $\mb{\Theta}\ts{h}$, the smoothed posterior of the ECM parameters was found by using the RTS smoother over the battery lifetime, which consisted of 19-20 points at the end of the chosen discharge cycles. With the smoothed estimates of $\mb{X}\ts{GP,t}$ and $\mb{P}\ts{GP,t}$, we also extrapolated the GP estimates to the correct point in time to validate future life predictions. At these points in the time axis, we used the standard GP predictive equations (\ref{eqn:GPposterior}) to retrieve the estimates of the functions over SOC and/or current, $I$.

\subsection{Results and Discussion}

\begin{figure*}
\centering
    \includegraphics[width=\textwidth]{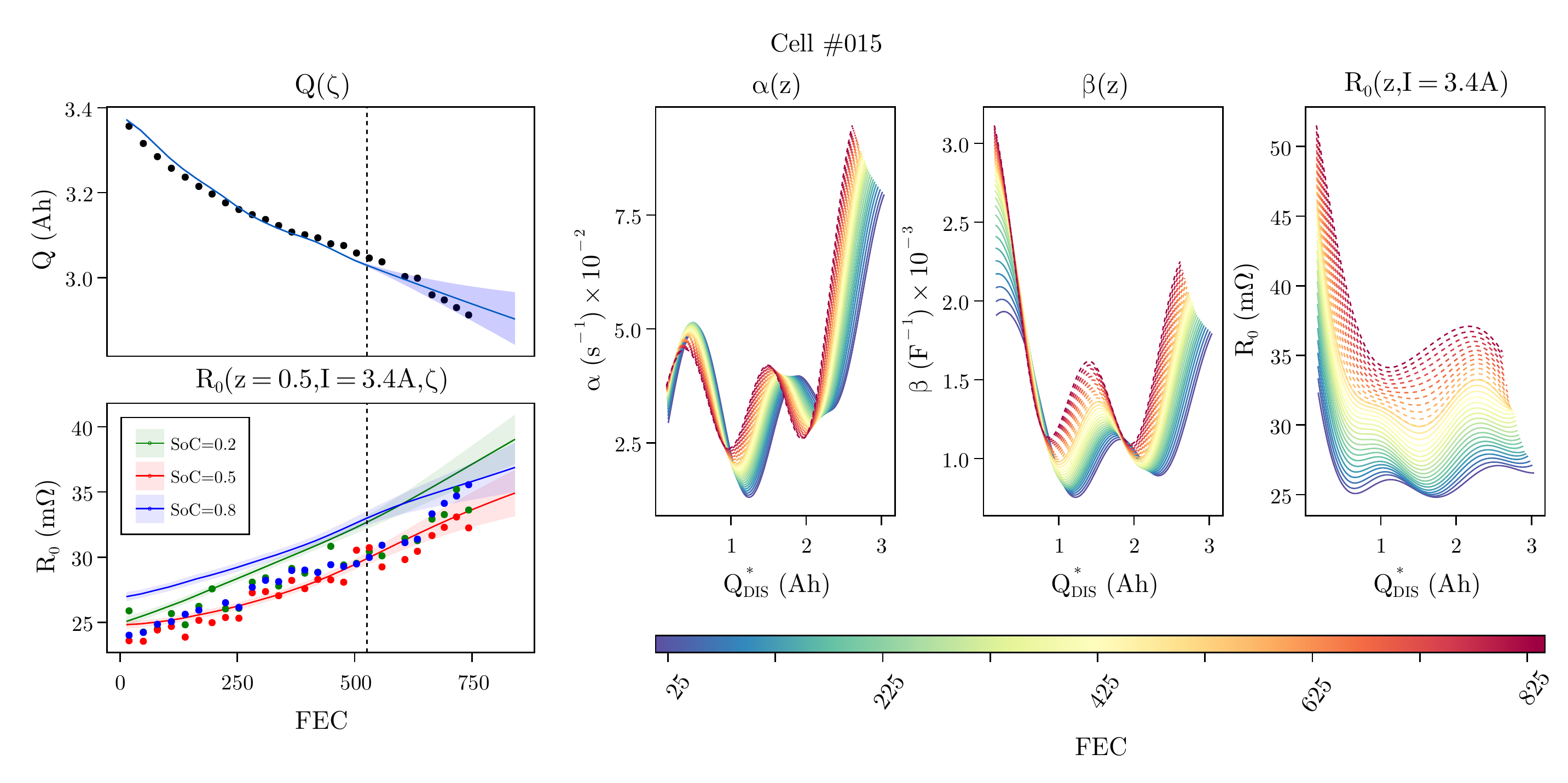}
    \caption{Projections of GP posteriors for $Q$, $R\ts{0}$, $\alpha$ and $\beta$. The validation data for $Q$ and $R\ts{0}$ (dots) show that the model accurately captures the evolution of SOH. The right three plots of $\alpha$, $\beta$ and $R\ts{0}$ as a function of discharge capacity show the strong dependency of ECM parameters on SOC (confidence bounds omitted for clarity). The colours indicate battery age expressed as full equivalent cycles (FEC), defined as the ratio of total Ampere-hour throughput to nominal  capacity, and the dashed lines denote extrapolated points in time equivalent to the $Q$ and $R\ts{0}$ points to the right of the vertical dashed lines.}    
    \label{fig:GP_all_data}
\end{figure*}

Fig.\ \ref{fig:GP_all_data} shows 1-D projections of the GP estimates for capacity $Q$, $R\ts{0}$, $\alpha$ and $\beta$ over battery lifetime for one of the cells chosen. As we used the summed NLML to determine hyperparameters using data from both cells in our sample, the results for the second cell are very similar (Appendix \ref{sec:other_cell}). In Fig.\  \ref{fig:GP_all_data} the GP posterior for the long-term smooth process is plotted, stripping out posterior of the noise process. The in-sample vs.\ out-of-sample division in each case is shown by the vertical black line. It is clear that the GP interpolated and extrapolated errors in capacity and $R\ts{0}$ are small and comparable  to the values retrieved during the independent checkup tests. The errors are shown in Table \ref{tab:performance}, with an average of \SI{0.016}{\ampere\hour} for capacity and \SI{1.7}{\milli\ohm} for internal resistance. The trend for $R\ts{0}$ is approximately linear over time and for capacity $Q$ we observe a reduced rate of degradation as the battery ages, until the final  checkup cycles where there is a possible increase in degradation rate. Extrapolation of the GP with the WV kernel is linear along the last known trajectory, which in this case is as accurate as the interpolation for both $Q$ and $R\ts{0}$.

The entire $R\ts{0}$ function shifts upwards as the cell ages, indicating a decrease in the reaction constant of either the anode or cathode. The degradation in capacity (top left Fig.\ \ref{fig:GP_all_data}) at early stages, measured during the first 20 checkup cycles, arguably has a slowing trend, which could imply diffusion-limited SEI growth on the anode \cite{Reniers2019}. If this is the dominant ageing mechanism, the increase in $R\ts{0}$ is most likely due to the anode reaction current (i.e.\ the product of electrode surface area and exchange current density) decreasing. In the last few checkup cycles, a slightly increased rate of capacity degradation is observed, although this is not reflected as an acceleration in the increase in $R\ts{0}$. The upward shift in $R\ts{0}$ with age is not evenly spread as a function of SOC (right subplot of Fig.\ \ref{fig:GP_all_data})---at high SOC (RHS of subplot), the curve shifts by approximately \SI{8}{\milli\ohm}, whereas low SOC (LHS of subplot), the shift is nearly \SI{20}{\milli\ohm}. The cells have a graphite anode with added silicon, and the latter often causes accelerated degradation of the electrode due to the large change in the volume of the silicon particles during charge and discharge, resulting in loss of active material \cite{kirkaldy_samieian_offer_marinescu_patel_2022}. As silicon participates most actively in the intercalation reactions at low SOC \cite{Ai2022AElectrodes}, it is consistent to see more substantial change in $R\ts{0}$ at low SOC due to mechanical degradation over cycling.

\begin{figure}
    \centering
    \includegraphics[width=\columnwidth]{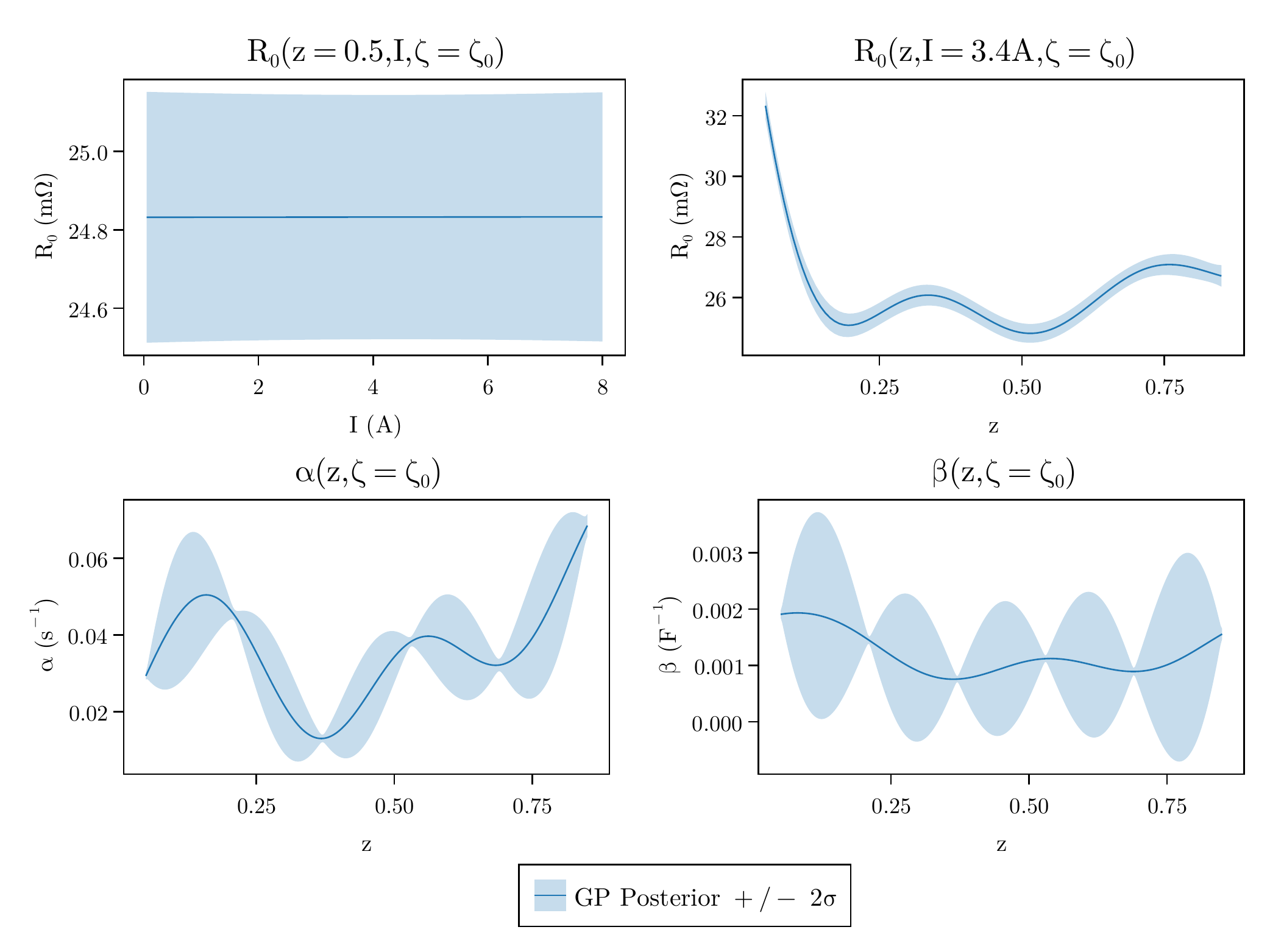}
    \caption{ECM Parameter function estimates at the first discharge cycle for the two cells. $R\ts{0}$ shows strong dependency on state of charge, but none on the applied current. $\alpha$ and $\beta$ also show SOC-dependence, with a short GP length scale implying high uncertainty in the process.}
    \label{fig:sc_functions}
\end{figure}

Each of the estimated functions are shown in more detail in Fig.\ \ref{fig:sc_functions}, showing the dependencies of parameters on SOC and current, including  beginning-of-life confidence bounds. Here we observe that $R\ts{0}$ is estimated to have no dependency on the applied current over the observed operating range, implying that linearisation of kinetics is probably reasonable here, but the SOC-dependency of $R\ts{0}$ has an estimated range of \SI{7}{\milli\ohm}. Similarly $\alpha$ and $\beta$ have substantial ranges over SOC, with an estimated GP length scale that is much shorter over SOC than for $R\ts{0}$. The short length scale is reflected in the rapid increase in predictive uncertainty in between the points at coordinates $\mb{u}$ that are propagated in state space. %
The evolution of the parallel RC-pair parameters over the battery lifetime is not as clear as it is for the series resistance. Overall, we observe less change in $\alpha$ and $\beta$ as a function of battery lifetime, but they both shift along the discharge capacity axis as the cell degrades.

\begin{figure}
\centering
    \includegraphics[width=\columnwidth]{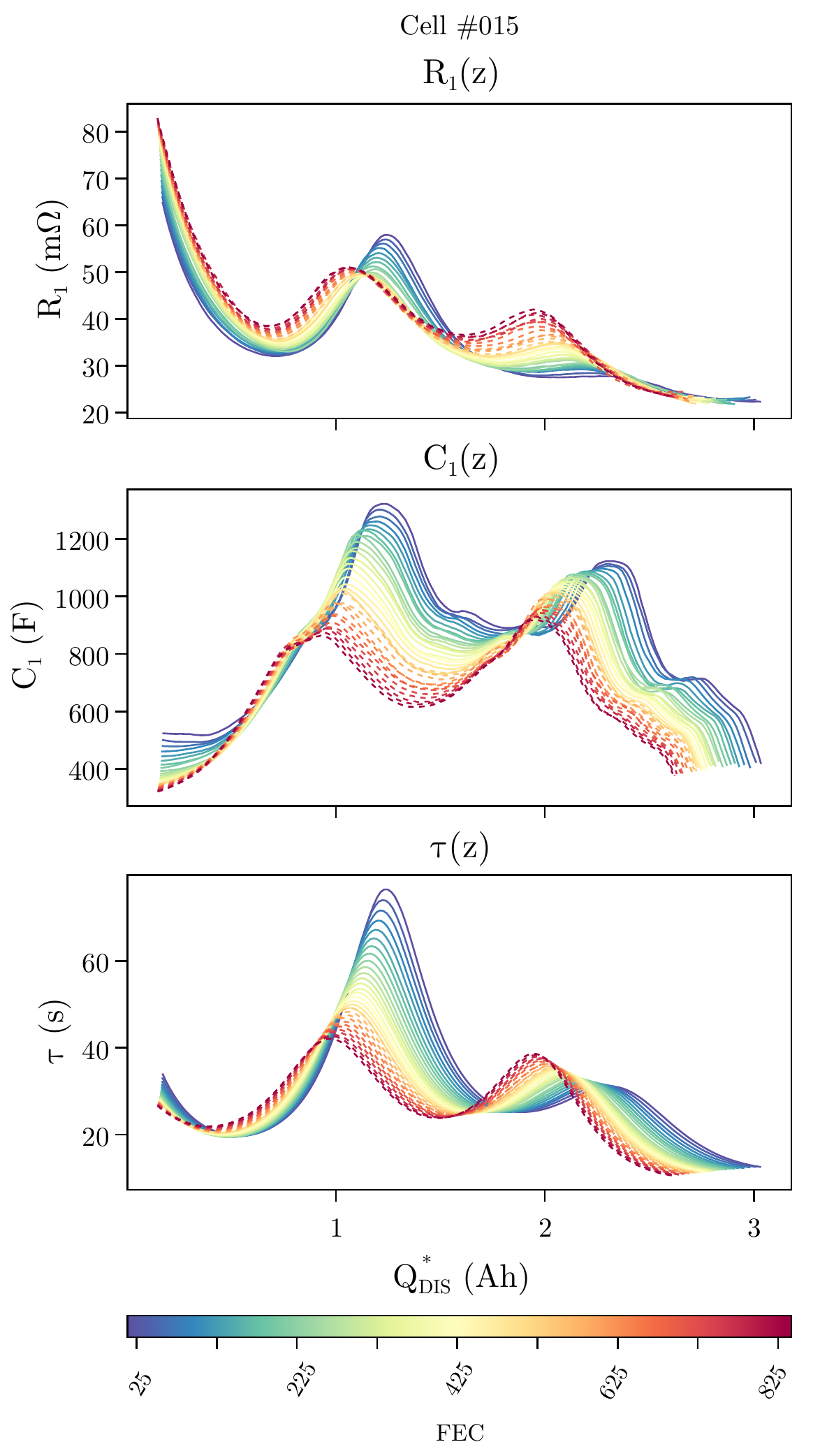}
    \caption{Conventional RC circuit parameterisation calculated by sampling from the ratios of the estimated GPs $\beta / \alpha$, $1/ \beta$  and $1/\alpha$,
    over battery lifetime.}
    \label{fig:GP_all_data_RC}
\end{figure}

The conventional RC parameterisation in terms of $R\ts{1}$ and $C\ts{1}$ may be retrieved from $\alpha$ and $\beta$, and this is shown in Fig.\ \ref{fig:GP_all_data_RC}. Because the nonlinear transformation of Gaussian variables (inversion in this case) does not yield a Gaussian, we show the 50\% percentile values for the two parameters over battery lifetime, which we sample from the posterior predictive ratios and inverses of $\alpha$ and $\beta$. The distributions are in fact heavily skewed because the GP for $\beta$ has significant probability mass around zero. Overall, we observe an average 11\% increase in $R\ts{1}$ and a 15\% decrease in $C\ts{1}$ over the battery lifetime for both cells. A direct physical interpretation of the shape of the functions $\alpha(z,\zeta)$ and $\beta(z,\zeta)$, or the equivalent $R\ts{1}$ and $C\ts{1}$, is not straightforward. As the first order RC-circuit only includes a single time constant, the GP estimate of the time constant is likely to be a weighted mixture of different processes at both electrodes, such as charge transfer and diffusion. Another possible interpretation for $\alpha$ and $\beta$ is the reaction relaxation time constant derived in Lin et al.\ \cite{Lin2022MultiscaleCells}, which describes the relaxation process of the inhomogeneity of SOC over the thickness of the electrode, originating from work by Newman and Tobias \cite{Newman1962TheoreticalElectrodes}. The time constant also shows very little drift over lifetime, although at high SOC there is some increase matched by a decrease at low SOC, Fig.\ \ref{fig:GP_all_data_RC}. This is the net effect of $R\ts{1}$ increasing by approximately 25\% on average and $C\ts{1}$ decreasing equally. Ultimately, the time constant from a single RC pair ECM is difficult to connect with physics-based models as it likely reflects an average of several processes.

\begin{table}[]
    \centering
    \input{Tables/performance}
    \caption{RMSE values for capacity and internal resistance for GP mean vs.\ checkup tests, split by cell and GP interpolation or extrapolation cases.}
    \label{tab:performance}
\end{table}

\section{Conclusions}

In this study, we have shown how Gaussian process regression may be used to estimate from data the dependency of battery equivalent circuit parameters on states and operating conditions. Recursive GP regression provides a computationally efficient framework for the coupled estimation of battery states and parameters that are functions of states. Incorporating GP regression into the state-parameter estimation problem has multiple advantages. Firstly, using a GP kernel function to describe the evolution of parameters over battery lifetime gives a flexible method to extrapolate parameters into the future---existing literature has either used random walks or simple deterministic models for this purpose. Secondly, by incorporating the operating-point dependency of each of the parameters, their estimates are more stable across battery lifetime in real-world scenarios where conditions vary. Imposing a GP prior also mitigates numerical ill-conditioning by acting as a regularisation mechanism in situation where parameters are not easily identifiable e.g.\ when estimating resistance with very low currents. Furthermore, the Bayesian framework used here provides estimates of parameter uncertainty,  which is a function of the model identifiability and the amount of data in the training set in the vicinity of the observed operating conditions and lifetime. Moreover, owing to its simplicity, the framework is chemistry and battery construction agnostic.

In contrast to physics-based models, ECMs  require minimal prior knowledge of battery parameters---only the open-circuit voltage curve is needed here. In our study, the thermal model was parameterised using a heat capacity value from literature. The thermal model, while not strictly necessary for the estimation of ECM parameters, gives another constraint on the dynamics and improves the identifiability of the system. 

For future work, the applicability of this method could be explored in the case where only partial charging or discharging is observed over lifetime, which would more realistically reflect battery usage. In this case, the assumption of the independence of the different stochastic processes controlling battery evolution could be relaxed. For example, by introducing a non-zero prior covariance between parameters, more easily identifiable parameters such as internal resistance could directly be mapped to remaining capacity, while taking into account the dependency of resistance on battery states and operating conditions.

\appendices

\section{GP Predictive distribution with input uncertainty}
\label{sec:z_unc}

In order to evaluate the battery dynamics in \eqref{eqn:KF_prop},  values for $\alpha$, $\beta$ and $R\ts{0}$ must be evaluated. Ordinarily, their predictive means and variances are given by (\ref{eqn:GPposterior}). However, because the state of charge, $z\ts{t}$, is itself a Gaussian random variable, the predictive distributions must be marginalised (averaged) over the distribution $z\ts{t}$ by performing the integral
\begin{equation}
        p(f(z\ts{t})) = \int p(f(z\ts{t})|z\ts{t})p(z\ts{t})\diff {z\ts{t}}
        \label{eqn:marginalise_z}
\end{equation}
at each time step, where $f(z\ts{t})$ represents the predictive distribution of $\alpha$, $\beta$ and $R\ts{0}$. In other words, the average GP prediction across all possible values of the probability distribution of the present SOC, $z\ts{t}$, is calculated. For an arbitrary kernel function $k$, this integral is usually intractable and can be approximated using e.g.\ Taylor expansions \cite{Girard2003GaussianForecasting}. However, for the SE kernel, (\ref{eqn:marginalise_z}) has an analytical solution given by Qui\~{n}onero-Candela et al.\ \cite{Quinonero-Candela2003PredictionForecasting}, with the mean and variance of $p(f(z\ts{t}))$ for scalar input $z$ being
\begin{equation}
\begin{aligned}
    \mathbb{E}[f(z)] & =  \mb{l}\ts{f,z}\tu{T}\mb{\delta}\\ 
    \mathbb{V}[f(z)] & = \sigma\ts{f,GP}^2(\mu\ts{z}) + \text{Tr}\left((\mb{K}^{-1} -\mb{k}\ts{f,z,\mb{u}}^{}\mb{k}\tu{T}\ts{f,z,\mb{u}})\mb{L}\right) \\
    & ~ ~ + \text{Tr}\left(\mb{\delta}\mb{\delta}\tu{T}(\mb{L}\ts{f,z}-\mb{l}\ts{f,z}^{}\mb{l}\ts{f,z}\tu{T})\right)
    \label{eqn:gp_pred_z_unc}
\end{aligned}
\end{equation}
where Tr denotes the matrix trace and $\mb{K} = \mb{K}\ts{\mb{f,uu}} + \text{diag}(\mathbf{P}\ts{f,t})$ for each of the GPs that are functions of $z\ts{t}$. The parameter $\delta$ is given by $\mb{K}^{-1}\mb{x}\ts{f,t}$, where $\mb{x}\ts{f,t}$ is the state vector for the GP at time t, and $\sigma\ts{GP}^2(\mu\ts{z})$ is the standard GP predictive variance evaluated at the mean of $z\ts{t}$ and $\mb{l}\ts{z}$, $\mb{L}\ts{z}$ are given by 
\begin{equation}
\begin{aligned}
    \mb{l}\ts{f,z} &= \frac{\sigma\ts{f,\mb{x}}^2}{\sqrt{\gamma\ts{f,z}\sigma^2\ts{z}+1}}\exp\left(-\frac{1}{2(\gamma\ts{f,z}^{-1}+\sigma^2\ts{z})}(\mu\ts{z}-\mb{{u\ts{f}}})^2\right) \\
    \mb{L}\ts{f,z} &=  \frac{\mb{k}\ts{f,z,u}^{}\mb{k}\ts{f,z,u}\tu{T}}{\sqrt{2\gamma\ts{f,z}\sigma^2\ts{z}+1}}\exp\left(\frac{2\gamma\ts{f,z}^2}{{2\gamma\ts{f,z}+\sigma^2\ts{z}}}(\mu\ts{z}-\mb{\overline{U}\ts{f}})^2\right).
    \label{eqn:Lls}
\end{aligned}
\end{equation}
Here, the vector $\mb{u}\ts{f}$ represents the coordinates of the discretisation points of $z$, and the element of matrix $\mb{\overline{U}\ts{f,ij}} =\frac{1}{2}(\mb{u}\ts{f,i} + \mb{u}\ts{f,j})$. The variable $\gamma\ts{f,z}$ is the inverse square of the length scale and $\sigma\ts{f,z}^2$ is the magnitude of GP $f$ over the SOC input $z$, and $\sigma\ts{z}^2$ is the variance of $z$. The vector $\mb{k}\ts{f,z,\mb{u}}$ is the kernel function $k\ts{SE}$ evaluated at $\mu\ts{z}$, i.e.\ the same as $\mb{l}\ts{f,z}$ where $\sigma^2\ts{z}=0$. %
In other words, uncertainty in $z\ts{t}$ affects both the mean and variance of the GP output, changing the weighting of the linear combination of points $x\ts{z}$ used to make a prediction and adding to the variance. As $\sigma\ts{x}^2 \rightarrow 0$, from equations (\ref{eqn:gp_pred_z_unc},\ref{eqn:Lls}) we recover the standard GP predictive equations.

\section{Predictive variances in EKF recursion}\label{sec:extra_vars}

The sparse matrices $\lambda$ in Eqns.\ (\ref{eqn:KF_prop}) and (\ref{eqn:KF_output}) arise from accounting for the extra uncertainties in state dynamics and predicted terminal voltage due to GP predictive variance. %
At each time-step, the GP predictive means and variances for $\alpha, \beta$ and $R\ts{0}$ are evaluated by Eqns.\ (\ref{eqn:gp_pred_z_unc},\ref{eqn:Lls}), which depend on $z\ts{t}$ and/or applied current.

The `extra variance' from GP interpolation or extrapolation in the battery state priors and terminal voltage predictions is encapsulated in the $\lambda$ terms in (\ref{eqn:KF_prop}) and (\ref{eqn:KF_output}). They are the result of the discretisation of the GPs over the inputs $z\ts{t}$ and $I\ts{t}$, which affects the joint probability distributions relating to the state dynamics and output predictions \cite{Aitio2021PredictingLearning,Sarkka2013}. In the general case, the conditional distributions are such that
\begin{subequations}
\begin{align}
\mathbf{x} &\sim \mathcal{N}(\mathbf{m},\mathbf{P}) \\    
\theta|\mathbf{x} &\sim \mathcal{N}(\mathbf{Hx},\mathbf{\Sigma}\ts{GP}) \\
\mathbf{y}|\mathbf{x},\theta &\sim \mathcal{N}(g(\mathbf{x},\theta),\Sigma\ts{n}),
\end{align}%
\end{subequations}
where the battery model parameter $\theta \in \{\alpha,\beta,R\ts{0}\}$ is a linear combination of the relevant GP states, which is then used in the state transition and observation function $g$ to yield the predictive prior states and outputs. In the EKF, we locally linearise $g$, giving the joint distributions of $\mathbf{x}$, $\theta$ and $\mathbf{y}$,
\begin{equation}
    \begin{bmatrix}
    \mathbf{x} \\
    \mathbf{\theta} \\
    \mathbf{y}
    \end{bmatrix} \sim \mathcal{N}(\mathbf{m^*},\mathbf{P^*}),
\end{equation}
where 
\begin{equation}
    \mathbf{m^*} = \begin{bmatrix}
    \mathbf{m} \\
    \mathbf{Hm} \\
    \mathbf{g(m,Hm)}
    \end{bmatrix},
\end{equation}
and 
\begin{equation}
\footnotesize
\mathbf{P^*} = \begin{bmatrix}
    \mathbf{P} & \mathbf{PH\tu{T}} & \mathbf{PH\tu{T}G\tu{T}} \\
    \mathbf{HP} & \mathbf{HPH\tu{T}} + \mathbf{\Sigma}\ts{GP} &  [\mathbf{HPH\tu{T}} + \mathbf{\Sigma}\ts{GP}]\mathbf{G\tu{T}}\\
    \mathbf{GHP\tu{T}} & \mathbf{G}[\mathbf{HPH\tu{T}} + \mathbf{\Sigma}\ts{GP}] & \mathbf{G}[\mathbf{HPH\tu{T}} + \mathbf{\Sigma}\ts{GP}]\mathbf{G\tu{T}} + \mathbf{\Sigma\ts{n}}
    \end{bmatrix},
    \label{eqn:P_joint}
\end{equation}
where $\mathbf{G}$ is the Jacobian of $g$. Because the model parameters that are discretised over $z\ts{t}$ and $I\ts{t}$ include $\alpha$, $\beta$ and $R\ts{0}$, $\mathbf{\Sigma\ts{GP}}$ will be very sparse---it will only have non-zero terms for the state transitions of $T$ and $V\ts{1}$ in (\ref{eqn:KF_prop}) and the terminal voltage prediction in (\ref{eqn:KF_output}). To simplify the calculation, we refactor the bottom right-hand term in (\ref{eqn:P_joint}) to obtain the $\lambda$ terms in (\ref{eqn:KF_prop},\ref{eqn:KF_output}), where the entries for each are given by
\begin{subequations}
\begin{align}
    \lambda\ts{G,t}[2,2] & = \sigma\ts{GP,\beta}^2\left(\frac{I\ts{t}(1-\exp(-\Delta t\mu\ts{\alpha,t}))}{\mu\ts{\alpha,t}}\right)^2+ \sigma\ts{GP,\alpha}^2V\ts{1,t}^2, \\
    \lambda\ts{G,t}[3,3] & =  \sigma\ts{GP,R\ts{0}}^2I\ts{t}^4\left(1-\exp(-\frac{\Delta t}{R\ts{c}C\ts{c}})\right)^2, \\
    \lambda\ts{H,t}[1,1] &=\sigma\ts{GP,R\ts{0}}^2I\ts{t}^2
\end{align}%
\end{subequations}

\section{Initialisation of WV process at beginning of life}\label{sec:WV_init}

The overall kernel describing each of of the four GPs in the circuit model (\ref{eqn:batt_dyn}) is separable over the operating point and battery lifetime and has a form
\begin{equation}
\small
\begin{aligned}
    k\ts{f} & = k\ts{\zeta}(\zeta,\zeta')k\ts{\mb{x}}(\mb{x},\mb{x}')  \\ 
    & = \sigma^2\ts{\zeta} \left(\frac{\min^3(\zeta,\zeta')}{3}+|\zeta-\zeta'|\frac{\min^2(\zeta,\zeta')}{2}\right) \times \\ & \quad~\sigma\ts{\mb{x}}^2\exp\left(-\frac{1}{2}\sum_\mb{x} \gamma\ts{\mb{x}}(\mb{x}-\mb{x}')^2\right),
\end{aligned}
\end{equation}
which variance at $\zeta=0$. To correct for this and to split the hyperparameter estimation problem in two (see Section \ref{sec:hyper_est_detail}), we reformulated the WV kernel so that its prior variance at beginning of life was equal to the $\sigma\ts{\mathbf{x}}^2$ for each of the GPs, where $\sigma\ts{\mathbf{x}}^2$ were estimated in the first phase of the estimation process. This is equivalent to shifting $\zeta$ to a non-zero starting point $\zeta_0$. To solve for the correct initial conditions in the recursive formulation, it is sufficient to solve (\ref{eqn:WV_sol}) for $\zeta_0$ in the top-left hand element, so that
\begin{equation}
    \frac{1}{3}\sigma\ts{\zeta}^2\zeta\ts{0}^3 = \sigma\ts{\mathbf{x}}^2
\end{equation}
for a given $\sigma\ts{\mathbf{x}}^2$ and $\sigma\ts{\zeta}^2$. This then yields the consistent $\mathbf{P}\ts{0}(\zeta_0)$ for each of the GPs estimated. 

\section{Results for data from second experimental cell}\label{sec:other_cell}

The second cell tester (number 009) has GP posteriors that show very similar patterns. This is due to the similarity in experimental conditions and the method of estimating GP hyperparameters, which was done by minimising the NLML using data from both cells, meaning that the GP hyperparameters we the same.

\begin{figure*}
\includegraphics[width=\textwidth]{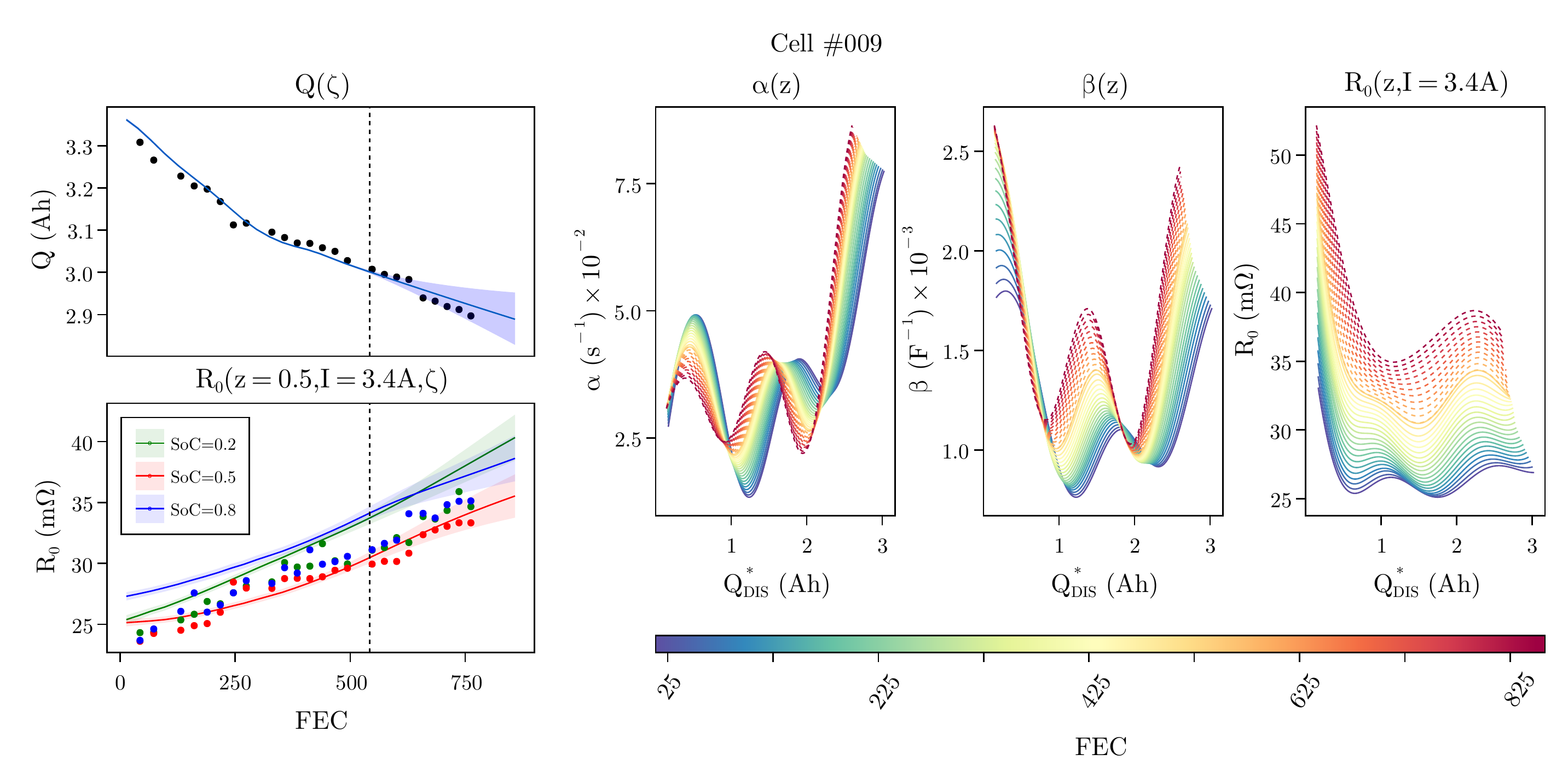}
\caption{GP posteriors for cell 009 in the experimental dataset, showing patterns consistent with cell 015.}
\end{figure*}

\begin{figure}
    \centering
    \includegraphics[width=\columnwidth]{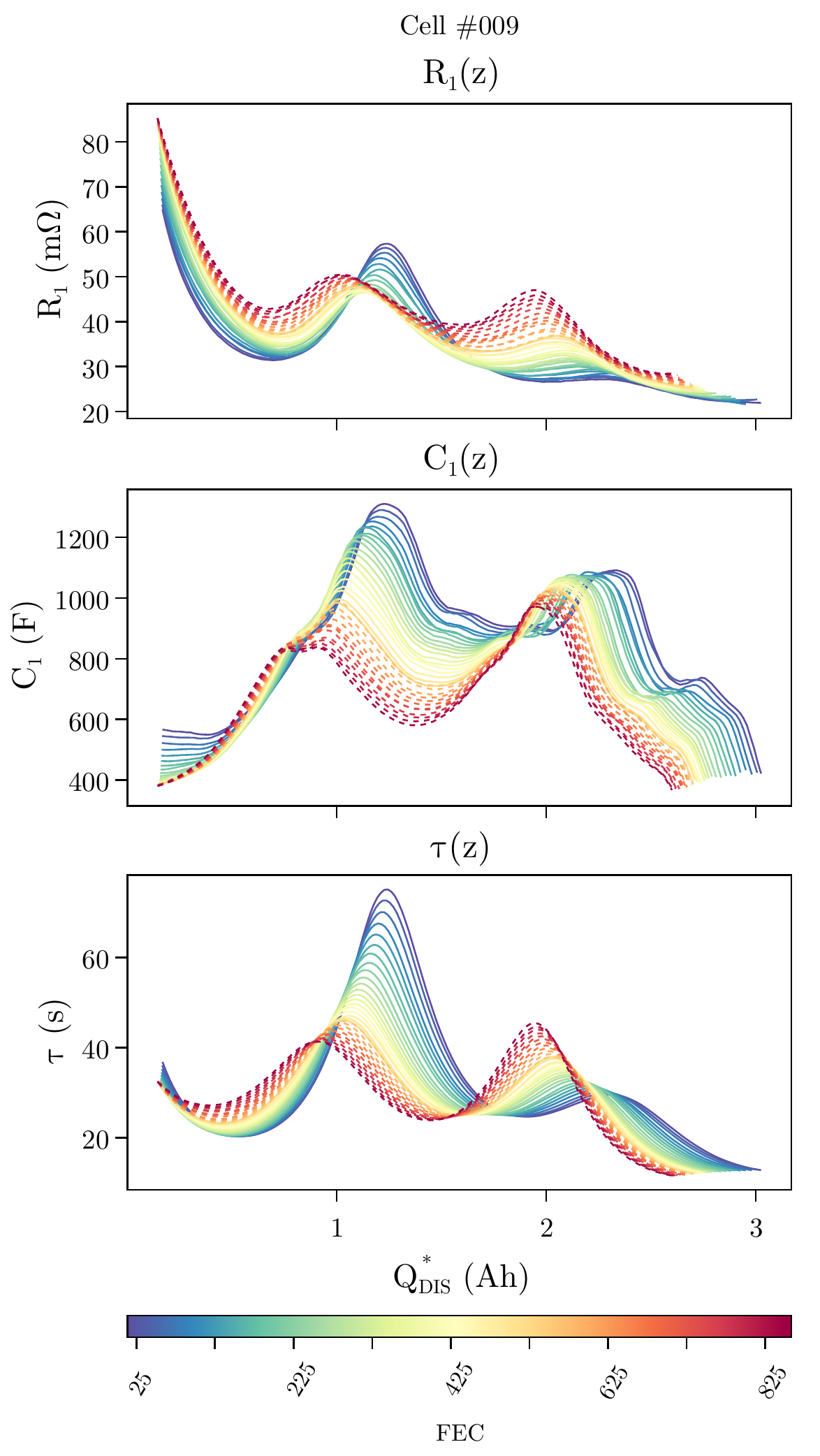}
    \caption{Cell 009 RC parameterisation.}
\end{figure}



\ifCLASSOPTIONcaptionsoff
  \newpage
\fi

\printbibliography

\end{document}

%% file: Tables/double_EKF_algo.tex
\begin{algorithmic}[1]
\State \textbf{Initialisation at $\zeta=\zeta\ts{0}$}
\State \hskip0.5em $\mb{x}^+\ts{GP}=\mb{x}^+\ts{GP,0}$ , $\mb{P}^+\ts{GP}=\mb{P}^+\ts{GP,0}$
\State \hskip0.5em $\phi\ts{t}=0$
\For{$\mathcal{D}\ts{s}$ $\in$ segments}
     \State \textbf{Initialisation at start of segment}
     \State \hskip0.5em $\mb{x}^+\ts{Batt}=\mb{x}\ts{Batt,0}$ , $\mb{P}^+\ts{Batt}=\mb{P}\ts{Batt,0}$ 
     \State \hskip0.5em $\mb{x}^+\ts{GP}=\exp(\mb{F}\Delta\zeta)\mb{x}^+\ts{GP}$ 
     \State \hskip0.5em $\mb{P}^+\ts{GP}=\exp(\mb{F}\Delta\zeta)\mb{P}^+\ts{GP}\exp(\mb{F}\Delta\zeta)\tu{T} + \mb{Q}\ts{GP}(\Delta\zeta)$ 
     \State \hskip0.5em $\mb{x}^+ = \begin{bmatrix}  \mb{x}^+\ts{Batt} &  \mb{x}^+\ts{GP} \end{bmatrix}\tu{T} $
     \State \hskip0.5em $\mb{P}^+ = \mathlarger{\oplus}\left(\mb{P}^+\ts{Batt} , \mb{P}^+\ts{GP}\right) $ 
    
     \For{I\textsubscript{t},V\textsubscript{t},T\textsubscript{t} $\in$ $\mathcal{D}\ts{s}$} 
     \State \textbf{Propagation}
     \State \hskip0.5em $\mb{x}^-\ts{t} = g(\mb{x}\ts{t-1}, I\ts{t-1}, T\ts{t-1})$ 
     \State \hskip0.5em $\mathbf{P}^-\ts{t} = \mathbf{G}\ts{t-1}\mathbf{P}^+\ts{t-1}\mathbf{G}\ts{t-1}\tu{T} + \mathbf{Q}\ts{t-1} +\lambda\ts{G,t-1}$
     \State \textbf{Observation and update}
     \State \hskip0.5em $\mb{e}\ts{t}  = [V\ts{t}~T\ts{t}]\tu{T} - h(\mb{x}\ts{t},I\ts{t})$ 
     \State \hskip0.5em $\mb{S}\ts{t} = \mb{H}\ts{t}\mb{P}\ts{t}^-\mb{H}\ts{t}\tu{T}+\mb{R}+\lambda\ts{H,t}$
     \State \hskip0.5em $\mb{K}\ts{t} = \mb{P}^-\ts{t}\mb{H}\ts{t}\tu{T}\mb{S}^{-1}\ts{t}$ 
     \State \hskip0.5em $\mb{x}\ts{t}^+ = \mb{x}^{-}\ts{t} + \mb{K}\ts{t}\mb{e}\ts{t}$ 
     \State \hskip0.5em $\mb{P}\ts{t}^+ = (\mb{I}-\mb{K}\ts{t}\mb{H}\ts{t})\tu{T} \mb{P}^-\ts{t} (\mb{I}-\mb{K}\ts{t}\mb{H}\ts{t}) + \mb{K}\ts{t}\mb{RK}\ts{t}\tu{T}$ 
     \State \hskip0.5em $\phi\ts{t} =\phi\ts{t} + \frac{1}{2} \mb{e}\ts{t}\tu{T}\mb{S}\ts{t}^{-1}\mb{e}\ts{t}+ \frac{1}{2}\log |2\pi \mb{S}\ts{t}|$ 
     \EndFor
\EndFor
\end{algorithmic}
    

%% file: Pics/Sim_IO_inkscape.pdf_tex
\begingroup%
  \makeatletter%
  \providecommand\color[2][]{%
    \errmessage{(Inkscape) Color is used for the text in Inkscape, but the package 'color.sty' is not loaded}%
    \renewcommand\color[2][]{}%
  }%
  \providecommand\transparent[1]{%
    \errmessage{(Inkscape) Transparency is used (non-zero) for the text in Inkscape, but the package 'transparent.sty' is not loaded}%
    \renewcommand\transparent[1]{}%
  }%
  \providecommand\rotatebox[2]{#2}%
  \newcommand*\fsize{\dimexpr\f@size pt\relax}%
  \newcommand*\lineheight[1]{\fontsize{\fsize}{#1\fsize}\selectfont}%
  \ifx\svgwidth\undefined%
    \setlength{\unitlength}{600bp}%
    \ifx\svgscale\undefined%
      \relax%
    \else%
      \setlength{\unitlength}{\unitlength * \real{\svgscale}}%
    \fi%
  \else%
    \setlength{\unitlength}{\svgwidth}%
  \fi%
  \global\let\svgwidth\undefined%
  \global\let\svgscale\undefined%
  \makeatother%
  \begin{picture}(1,0.75)%
    \lineheight{1}%
    \setlength\tabcolsep{0pt}%
    \put(0,0){\includegraphics[width=\unitlength,page=1]{Sim_IO_inkscape.pdf}}%
  \end{picture}%
\endgroup%

%% file: Tables/sim_parameters.tex
\small
    \centering
    \begin{tabular}{ccccc}
    \toprule
    Parameter & Description & Value  &Units & GPR RMSE (\%) \\ 
    \midrule
         $\alpha(z)$ & Inverse RC pair time constant & $0.015-0.09\left(0.05-z\right)^3$ & \SI{}{\per\second} & 3.1 \\
         $\beta(z)$ & Inverse RC pair capacitance & $0.002\left(1-(z-0.5)^2\right)$ & \SI{}{\per\farad} & 0.73 \\
         $R\ts{0}(z,I)$ & Series resistor & $0.05\sinh^{-1}\left(|I|\right)/|I|+0.04\left(z-1\right)^2$ & \SI{}{\ohm} & 0.97 \\
         $Q^{-1}$ & Inverse cell capacity & 1.2 & \SI{}{\per\amperehour} & 0.15\\
         $R\ts{c}$ & Thermal resistance & 5.5 & \SI{}{\kelvin\per\watt} & Given \\
         $C\ts{c}$ & Heat capacity & 15.7 & \SI{}{\joule\per\kelvin} & Given \\
         $V\ts{0}$ & Open-circuit potential & $3.64 + 0.55x - 0.72x^2 + 0.75x^3$ &  \SI{}{\volt} &  Given \\
         $\sigma\ts{n,V}$ & Voltage measurement noise st. dev. & 0.005 & \SI{}{\volt} & 0.4$\dagger$\\
         $\sigma\ts{n,T}$ & Temperature measurement noise st. dev. & 0.1 & \SI{}{\kelvin} & 2.8$\dagger$ \\
         
    \bottomrule \\
    \end{tabular}
\caption{Simulation parameters and errors in estimates. The error in GPR estimates after a single discharge cycle is $<$4\% for each circuit parameter, and is not strongly affect by adding more data. $\dagger$ these are a subset of the GP hyperparameters.}

%% file: Tables/performance.tex
  \begin{tabular}{cccccc}
    \toprule
    \textbf{Cell} & \textbf{Case} & \textbf{Q (Ah)} & \multicolumn{3}{c}{\textbf{R$_0$(z, I = 3.4 A) (m$\Omega$)}}  \\
     &  &  & \texttt{z=0.2} & \texttt{z=0.5} & \texttt{z=0.8} \\\midrule
    9 & Interpolated & 0.018 & 2.0 & 0.92 & 2.7 \\
     & Extrapolated & 0.014 & 2.2 & 0.98 & 1.8 \\
    15 & Interpolated & 0.017 & 2.0 & 0.83 & 2.8 \\
     & Extrapolated & 0.016 & 1.9 & 1.5 & 1.7 \\\bottomrule
  \end{tabular}